\documentclass[journal=acsnano,manuscript=communication,layout=twocolumn]{achemso}

\usepackage{chemformula} 
\usepackage[T1]{fontenc} 

\usepackage{tikz}

\usepackage{etoolbox}
\patchcmd{\section}
  {\centering}
  {\raggedright}
  {}
  {}
\patchcmd{\subsection}
  {\centering}
  {\raggedright}
  {}
  {}
\usepackage{upgreek}
\usepackage{tcolorbox}
\usepackage[sort&compress]{natbib}
\usepackage{soul}
\usepackage{graphicx}
\usepackage{epstopdf}
\usepackage{verbatim}
\usepackage{float}
\usepackage{sidecap}
\sidecaptionvpos{figure}{c}
\usepackage{lipsum}
\usepackage{setspace}
\usepackage{dcolumn}
\usepackage{bm}
\usepackage{amssymb}
\usepackage{amsmath}	
\usepackage{mathtools}
\usepackage{color}
\usepackage{multirow}
\usepackage{lipsum}
\usepackage[colorlinks=true, urlcolor=blue, pdfborder={0 0 0}]{hyperref}
\hypersetup{
linkcolor=blue,
citecolor=blue
}

\newcommand{\kdotp}{$\textbf{k}\cdot\textbf{p}$ }

\newcommand{\GaBiAs}{GaBi$_{x}$As$_{1-x}$ }
\newcommand{\GaBiAsGaAs}{GaBi$_{x}$As$_{1-x}$/GaAs }
\newcommand{\GaAsGaBiAsGaAs}{GaAs/GaBi$_{x}$As$_{1-x}$/GaAs }

\let\oldmaketitle\maketitle
\let\maketitle\relax 

\author{Muhammad Usman}
\affiliation{School of Computing and Information Systems, Melbourne School of Engineering, The University of Melbourne, Parkville, 3010, Victoria, Australia.}
\alsoaffiliation{School of Physics, The University of Melbourne, Parkville, 3010, Victoria, Australia.}
\email{musman@unimelb.edu.au}

\title[An \textsf{achemso} demo]
  {Tunable band-gap and isotropic light absorption from bismuth-containing GaAs core$-$shell and multi$-$shell nanowires}

\keywords{Nanowire, Photonics Devices, Strain, Optical Transitions, Charge Confinements}

\begin{document}





\twocolumn[
\begin{@twocolumnfalse}
\oldmaketitle
\begin{abstract}
Semiconductor core$-$shell nanowires based on the GaAs substrate are building blocks of many photonic, photovoltaic and electronic devices, thanks to the associated direct band-gap and the highly tunable optoelectronic properties. The selection of a suitable material system is crucial for custom designed nanowires tailored for optimised device performance. The bismuth containing GaAs materials are an imminent class of semiconductors which not only enable an exquisite control over the alloy strain and electronic structure but also offer the possibility to suppress internal loss mechanisms in photonic devices. Whilst the experimental efforts to incorporate \GaBiAs alloys in the nanowire active region are still in primitive stage, the theoretical understanding of the optoelectronic properties of such nanowires is only rudimentary. This work elucidates and quantifies the role of nanowire physical attributes such as its geometry parameters and bismuth incorporation in designing light absorption wavelength and polarisation response. Based on multi-million atom tight-binding simulations of the \GaBiAsGaAs core$-$shell and \GaAsGaBiAsGaAs multi$-$shell nanowires, our results predict a large tuning of the absorption wavelength, ranging from 0.9 $\mu$m to 1.6 $\mu$m, which can be controlled by engineering either Bi composition or nanowire diameter. The analysis of the strain profiles indicates a tensile character leading to significant light-hole mixing in the valence band states. This offers a possibility to achieve polarisation-insensitive light interaction, which is desirable for several photonic devices involving amplification and modulation of light. Furthermore, at low Bi compositions, the carrier confinement is quasi type-II, which further broadens the suitability of these nanowires for myriad applications demanding large carrier separations. The presented results provide a systematic and comprehensive understanding of the \GaBiAs nanowire properties and highlight new possibilities for future technologies in photonics, quantum optics and solar energy harvesting.
\end{abstract}
\end{@twocolumnfalse}
]

\noindent
Semiconductor nanowires are versatile nanostructures which provide access to a unique set of electronic and optical characteristics suitable for many photonic applications~\cite{Dasgupta_advm_2014, Zhuang_Advm_2011}. The recent advancements in the growth techniques have allowed an exquisite control of nanowire morphologies, leading to engineered properties tailored for device functionalities. This has opened up tremendous opportunities for application of nanowires in the realization of a wide range of high-performance optoelectronic devices, including solar cells, photodetectors, light sensors, optical modulators, optical amplifiers and solid-state light sources~\cite{Yao_NNano_2013, Tomioka_Nature_2012, Chen_NatureP_2011, Wang_Nanoletters_2003, Yan_Nature_Photo_2009, Eaton_Nature_Review_2016}. More recently, there are proposals for novel applications of nanowires in cross-disciplinary fields such as medical imaging and solar to fuel conversion. Consequently, in the last few years, semiconductor nanowires have been a topic of major research efforts on both experimental and theoretical fronts, where the focus has been to optimise nanowire parameters such as its geometry dimensions and composition to enable directed applications in next generation technologies.

For nanowire growth and design, a large number of material systems have been investigated in the literature including III-Vs (such as GaAs, GaNAs, InGaP, GaPAs)~\cite{Yao_NNano_2013, Tomioka_Nature_2012, Chen_NatureP_2011, Wang_Nanoletters_2003, Yan_Nature_Photo_2009, Eaton_Nature_Review_2016}, SiGe~\cite{Cartoixa_Nanolett_2017}, and ZnO~\cite{Zhao_Nanolett_2020}, among several others. The nanowires made up of GaAs based III-V materials have been of particular interest due to highly promising optoelectronic characteristics, possibilities of dislocation free epitaxial growth, and monolithic integration with a large number of substrates including technologically relevant silicon~\cite{Tomioka_Nature_2012, Dimakis_NanoLett_2014} and graphene~\cite{Munshi_NanoLett_2012}. One particular class of nanowires is known as core$-$shell nanowires, where the core and shell regions are made up of different III-V materials such as GaAs and InGaAs~\cite{Royo_JPDAP_2017}. There has also been studies for designing nanowires with multi$-$shell structures where each shell is grown with a different alloy composition to allow engineered carrier confinement and light interactions~\cite{Balagula_SciRep_2020}. The core$-$shell nanowires have been shown to offer excellent tuning of optoelectronic properties because the strain and electronic confinement can be controlled by designing the size of the core and shell regions as well as the alloy compositions. 

The nanowires formed by the conventional GaAs alloy materials such as InGaAs, InGaP, and AlGaAs have been around for many years and their optoelectronic properties are well studied in the literature~\cite{Kim_Nanoletters_2006, Ren_Adv_Mat_2014, Ma_Nano_Lett_2014, Dai_NanoLett_2014, Dimakis_NanoLett_2014}. However, very recently, the focus has been shifted towards incorporating highly-mismatched bismuth containing GaAs materials such as \GaBiAs in the active region of core$-$shell nanowires~\cite{Ishikawa_Nanoletters_2015, Steele_SciRep_2016, Zelewski_APL_2016, Wang_APE_2016, Oliva_arXiv_2019, Zhang_NanoLett_2019, Matsuda_NanoLett_2019, Matsuda_JAP_2019}. The \GaBiAs materials which are formed by adding dilute concentration of bismuth (Bi) in the GaAs material offer unique electronic properties which are not readily accessible from traditional III-V alloys. It has been shown that the band-gap energy of the \GaBiAs material decreases dramatically with increasing Bi composition~\cite{Janotti_PRB_2002, Zhang_PRB_2005, Usman_PRB_2011, Bismuth_containing_compounds_2013, Broderick_bismide_chapter_2017, Fluegel_PRL_2006}. This implies that the band-gap energy can be tuned to a wide spectral range, encompassing near, mid and far infra-red bands (1-10 $\mu$m). The spectroscopic ellipsometry study has shown that the refractive index for the \GaBiAs material increases with Bi fraction and is slightly larger than that of the GaAs material~\cite{Tumenas_PSSc_2012}. Another highly promising property of the \GaBiAs material system originates from very large spin-orbit coupling associated with the GaBi material ($\approx$2.3 eV), which leads to a spin-orbit splitting energy being greater than the band-gap energy in the telecommunication spectral range (1550 nm) $-$ a property which is not available from other GaAs based materials. The spin-orbit splitting energy larger than the band-gap energy is highly desirable for the suppression of internal loss mechanisms such as inter-valence band absorption (IVBA) and CHSH Auger recombination processes which plague the efficiency of many photonic devices~\cite{Batool_JAP_2012, Usman_PRB_2011, Phillips_IEEEJSTQE_1999, Sweeney_PSSB_1999, Broderick_SST_2012}. Therefore, a systematic and comprehensive understanding of the \GaBiAs core$-$shell and multi$-$shell nanowires such as presented in this work is expected to play an important role towards exploiting the unique characteristics of this imminent material system for the next generation photonic devices.  
 
The interest in the \GaBiAs nanowires is at its primitive stage and the first few experimental studies have been reported in the literature during the last three to four years~\cite{Ishikawa_Nanoletters_2015, Steele_SciRep_2016, Zelewski_APL_2016, Wang_APE_2016, Oliva_arXiv_2019, Zhang_NanoLett_2019, Matsuda_NanoLett_2019, Matsuda_JAP_2019}. Theoretically, \GaBiAs core$-$shell and multi$-$shell nanowires have not been studied in much detail. Recently, \GaBiAsGaAs nanowires were theoretically investigated where the core region was made up of the \GaBiAs material surrounded by a GaAs shell region~\cite{Usman_Nanoscale_2019}. Here, in this work the focus is on \GaBiAs nanowires where the core region is made up of GaAs region and the \GaBiAs alloys is present in the shell region. Such nanowires are highly relevant to many experimental studies on the III-V core$-$shell nanowires where a tertiary or quaternary alloy is typically present in the shell region and the core region is a binary material. As the experimental research on \GaBiAs nanowires is rapidly advancing, there are several open questions regarding incorporation of Bi in the nanowire active region, such as the band-gap dependence on the Bi fraction, the charge carrier confinements and the sensitivity to the light polarisation. The answers to these questions are crucial to target nanowire parameters tailored for photonic applications. This work aims to provide a timely guidance for future experiments and will contribute significantly in the advancement of an emerging area of research.

This work is based on the state-of-the-art atomistic tight-binding simulations, which have been carried out on 4 to 8 million atoms to understand the optoelectronic properties of \GaBiAs nanowires with realistic geometry parameters. The accuracy of the tight-binding model was verified against the available experimental data sets in a number published studies including on band-gap variation of the \GaBiAs bulk material as a function of Bi composition with~\cite{Usman_PRB_2013} and without~\cite{Usman_PRB_2011} strain, and GaBiAs/GaAs quantum well structures~\cite{Usman_PRA_2018}. Subsequent studies based on DFT model~\cite{Kudrawiec_JAP_2014, Polak_SST_2015, Bannow_PRB_2016} and experimental measurements~\cite{Donmez_SST_2015, Balanta_JoL_2017, Zhang_JAP_2018, Dybala_APL_2017, Collar_AIPA_2017, Usman_PRB_2011, Usman_PRB_2013, Broderick_PRB_2014} demonstrated a good agreement with the simulated results, which again affirmed the high-level accuracy of our atomistic techniques. We also note that in contrast to the simplified models used in the literature such band anti-crossing method, effective-mass and \kdotp models~\cite{Broderick_SST_2012, Broderick_SST_2013, Broderick_SST_2013, Broderick_SST_2015} which have been quite successful to qualitatively model \GaBiAs based photonic devices, the atomistic tight-binding simulations employed here explicitly represent the nanowire geometries with atomic resolution. This allows to incorporate the impact of additional effects such as alloy randomness and interface roughness, which have been shown to play an important role in the investigation of electronic and optical properties of \GaBiAs materials~\cite{Usman_PRB_2013, Usman_APL_2014} and more generally for several other nanostructures~\cite{Klimeck_IEEETED_2007_2, Ahmed2009}. Therefore, this work is expected to provide a highly reliable quantitative understanding of the optoelectronic properties of the investigated \GaBiAs core$-$shell nanowires, which will be directly relevant to the ongoing experimental efforts.

\begin{figure*}[t]
\includegraphics[scale=0.195]{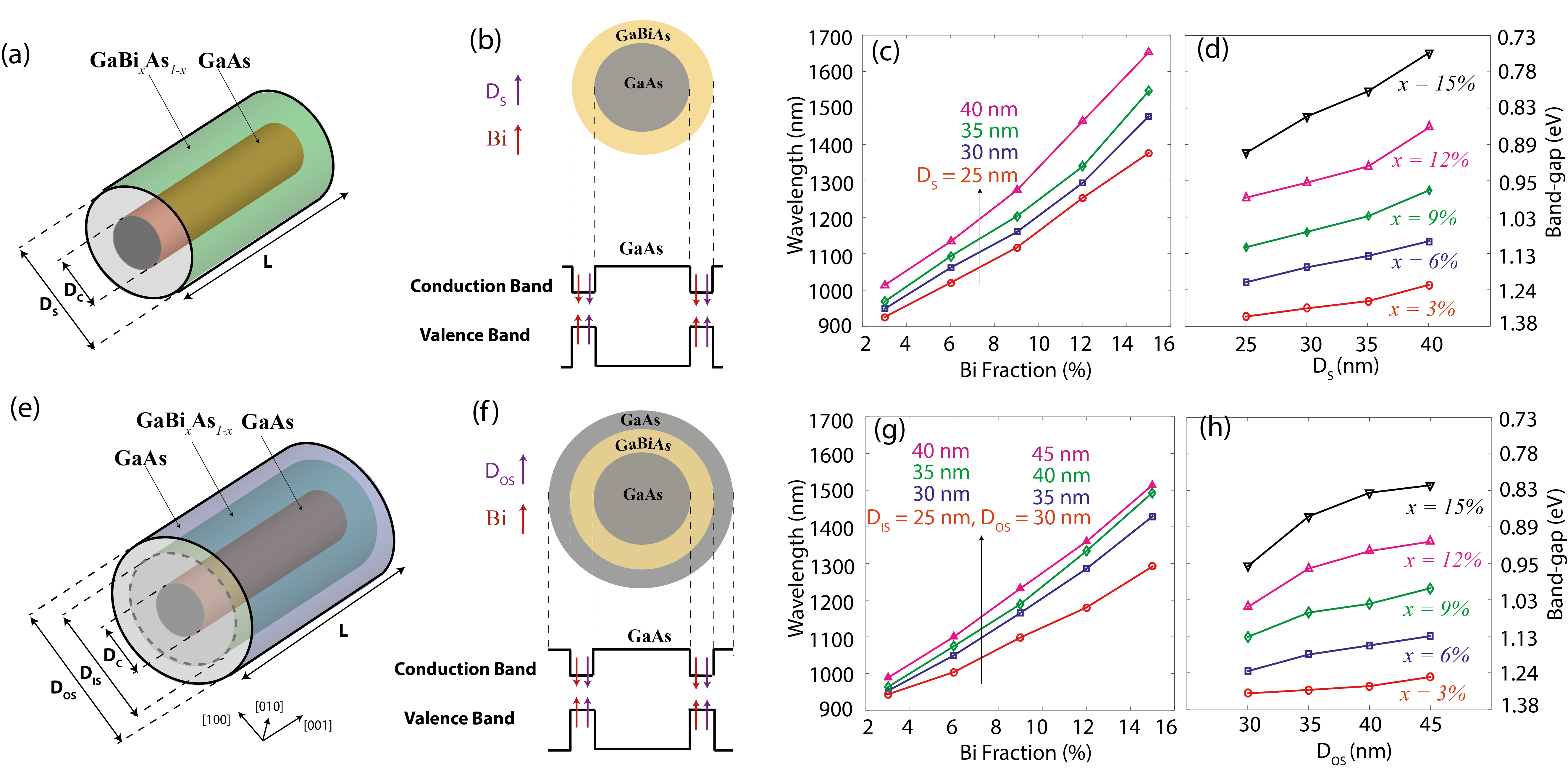}
\caption{\textbf{Band-gap tuning.} \textbf{(a)} Schematic illustration of the investigated \GaBiAsGaAs core$-$shell nanowire is shown. The nanowire consists of a \GaBiAs shell with diameter $\rm D_S$ and a GaAs core region with diameter $\rm D_C$. The length of the nanowire is L along the [001] direction. We note that the shell diameter of nanowire is same as the overall diameter of the nanowire structure. \textbf{(b)} Band edge diagram is schematically shown to indicate the effect of increasing Bi fraction and nanowire diameter on the conduction band and valence band edge energies. The direction of the arrows indicates the increase/decrease in the band edge energies when the Bi fraction or nanowire diameter is increased. \textbf{(c)} The ground-state inter-band absorption band-gap energy/wavelength is plotted as a function of the Bi fraction ($x$) in the nanowire shell region for various geometry parameters. \textbf{(d)} The plots of band-gap energy/wavelength are shown as a function of the nanowire shell diameter ($\rm D_S$) for various values of Bi fraction. \textbf{(e)} Schematic diagram is shown for the investigated \GaAsGaBiAsGaAs core$-$shell nanowire. The core region is made up of GaAs material with diameter $\rm D_C$. The nanowire consists of two shell regions, an inner \GaBiAs shell with diameter $\rm D_{IS}$ and an outer GaAs shell with diameter $\rm D_{OS}$. \textbf{(f)} Same plot as in (b) but for a \GaAsGaBiAsGaAs nanowire. \textbf{(g)} The plots of the ground-state inter-band absorption band-gap energy/wavelength are shown for various nanowire geometry parameters when the Bi fraction is varied from 0 to 15\%. \textbf{(h)} The plots of band-gap energy/wavelength are shown as a function of the nanowire outer shell diameter ($\rm D_{OS}$) for various values of Bi fraction.}
\label{fig:Fig1}
\end{figure*}     

The reported results have uncovered novel properties of the \GaBiAsGaAs nanowires which further motivates interest in this material system and highlight opportunities for a wide range of applications. We show that the inter-band absorption wavelength is highly sensitive to the nanowire diameter and Bi fraction of the shell region, and can be tuned from 0.9 $\mu$m to 1.6 $\mu$m. This provides a two-way knob to control light absorption/emission energies. Another interesting property of the \GaBiAsGaAs nanowires reported in this work is the transition from quasi type-II carrier confinement to type-I carrier confinement, which occurs at low Bi fractions ($\approx$3\%). This property is useful for designing a variety of devices where large carrier separation or strong light absorption is the parameter of interest. Finally, our results demonstrate that the \GaBiAsGaAs nanowires provide a strong light-hole absorption, which couples to TM polarisation mode. This is in contrast to usual III-V nanostructures where typically TE mode absorption is dominant. We also show the possibility of isotropic light polarisation at 1.55 $\mu$m wavelength which is important for many optical devices such as semiconductor optical amplifiers and modulators. Overall, our results indicate that \GaBiAsGaAs nanowires can be custom designed to target desired properties and hence are expected to find myriad applications in nanophotonic and nanoelectronic technologies.  
\\ \\
\noindent
\textbf{Results and Discussions}
\\ \\
\noindent
In this work, we have investigated two types of core$-$shell nanowires as shown by the schematic diagrams in Figure~\ref{fig:Fig1}. Both types of nanowire consist of a GaAs core region with a core diameter ($\rm D_C$) of 20 nm. The \GaBiAsGaAs nanowire in Figure~\ref{fig:Fig1} (a) has a single outer shell made up of \GaBiAs material with a shell diameter ($\rm D_S$) varying from 25 nm to 40 nm. The Bi composition of the shell region is varied from 3\% to 15\%. The second type of nanowire \GaAsGaBiAsGaAs studied in this work consists of two shell regions: an inner shell made up of \GaBiAs material with diameter $\rm D_{IS}$ and an outer shell made up of GaAs material with diameter $\rm D_{OS}$. Such nanowire structures are also referred to as quantum well nanowires in the literature~\cite{Qian_NM_2008} and provide an extra degree of freedom to engineer nanowire properties. The diameter $\rm D_{IS}$ is varied from 25 nm to 40 nm as for the previous case of \GaBiAsGaAs nanowires. The thickness of the outer shell in all cases is selected as 5 nm, therefore $\rm D_{OS}$ varies from 30 nm to 45 nm. The nanowires are constructed atomistically, where the \GaBiAs shell region contains Bi atoms randomly replacing As atoms. The length of nanowires is selected as 80 nm, however the previous study has shown that the electronic band-gap energy is relatively independent of the nanowire length. The largest dimension of the \GaBiAsGaAs and \GaAsGaBiAsGaAs nanowires consists of roughly 6.3 million and 8 million atoms respectively, in the simulation domain. The number of Bi atoms in the active region for 15\% Bi composition are around 0.47 million. The nanowires structures are relaxed by valence force field (VFF) model and the internal strain is computed from the relaxed atom positions. The electronic structure is computed by solving a ten-band \textit{sp$^3$s$^*$} tight-binding Hamiltonian which explicitly includes spin-orbit coupling. The lowest electron and the highest hole energies are computed at the $\Gamma$ point. Further details of the methods are provided in the Methods section.     
\\ \\
\noindent
\textbf{Band-gap Tuning}
\\
Figure~\ref{fig:Fig1} (a) shows the schematic diagram of the studied \GaBiAsGaAs core$-$shell nanowires. The band-gap energy of the nanowire is investigated by varying two parameters: the Bi fraction ($x$) in the \GaBiAs shell region and the diameter ($\rm D_S$) of the shell region. The changes in the conduction and valence band edges are schematically indicated in Figure~\ref{fig:Fig1} (b) when $x$ and $\rm D_S$ are increased. The increase in the Bi fraction reduces the conduction band edge energy and increases the valence band edge energy, in agreement with the previous studies on \GaBiAs bulk and quantum well systems~\cite{Usman_PRB_2011, Usman_PRB_2013, Usman_PRA_2018}. We find the same trend when the shell diameter of the nanowire is increased. Consequently, the band-gap energy decreases as a function of both Bi fraction and shell diameter. The band-gap dependence on the Bi fraction is plotted in Figure~\ref{fig:Fig1} (c) for various shell diameters. Our calculations show that by increasing the Bi fraction from 3\% to 15\%, the band-gap energy can be tuned from $\approx$1.38 eV to $\approx$0.78 eV (corresponding to a change of 900 nm to 1600 nm in the optical absorption wavelength). Likewise, a reduction in the band-gap energy is computed with an increase in shell diameter, which is stronger at larger Bi compositions. We compute a red shift of about 36 meV and 40 meV per \% increase in the Bi fraction at $\rm D_S$ = 25 nm and 40 nm, respectively. Figure~\ref{fig:Fig1} (d) shows the dependence of the band-gap energy on the nanowire shell diameter when the Bi faction is kept constant. The decrease(increase) in the band-gap energy(wavelength) is stronger at 15\% Bi fraction when compared to the 3\% Bi composition. Overall, we find that the band-gap energy is highly sensitive to both Bi composition and the nanowire diameter, and can be tuned in the telecommunication spectra. 

\begin{figure*}[t]
\includegraphics[scale=0.395]{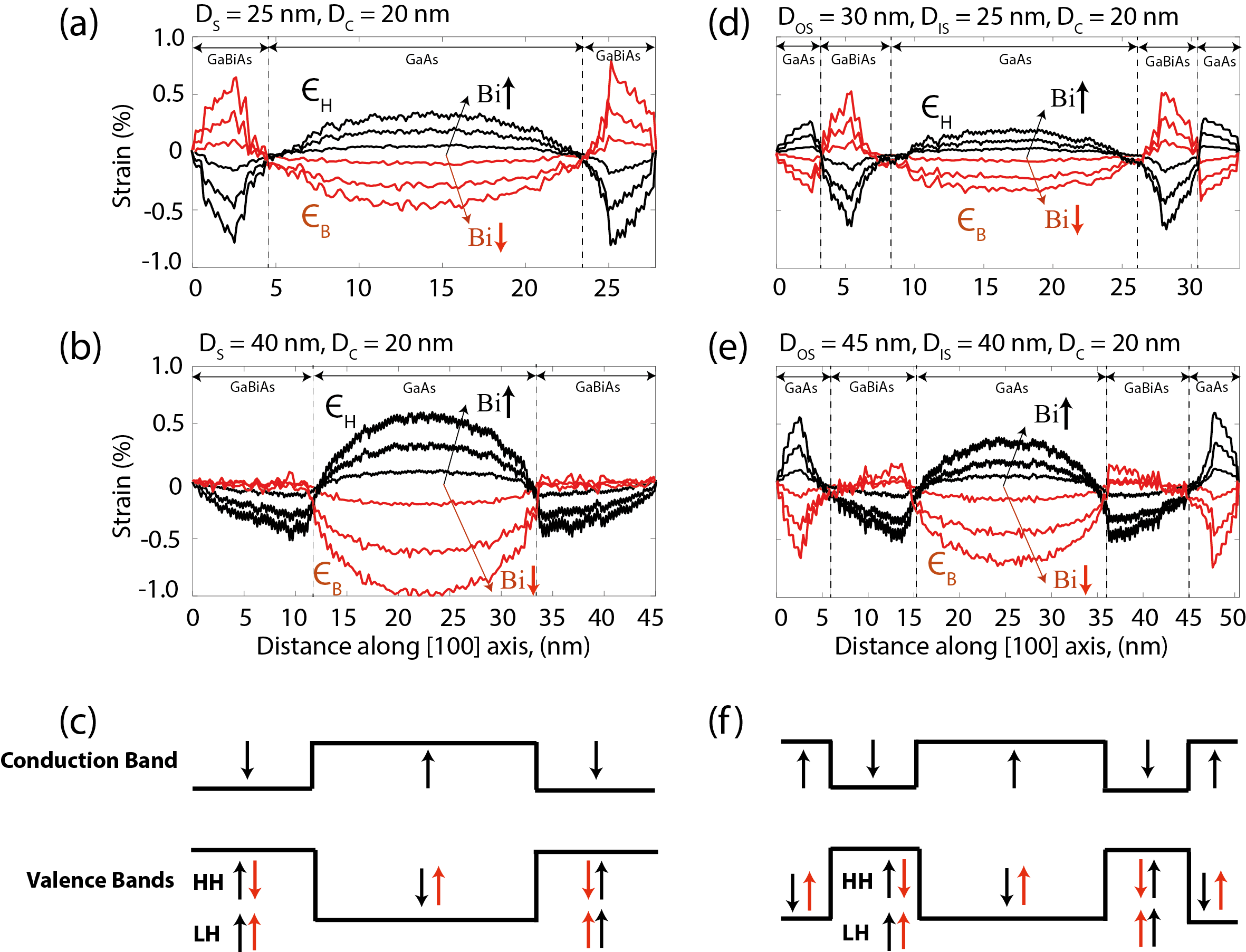}
\caption{\textbf{Strain profile analysis.} \textbf{(a)} The plots of the hydrostatic ($\epsilon_H = \epsilon_{xx} + \epsilon_{yy} + \epsilon_{zz} $) and biaxial ($\epsilon_B = \epsilon_{xx} + \epsilon_{yy} - 2\epsilon_{zz}$) strain components are plotted with black and red colors respectively as a function of the distance along the [001] axis through the center of the nanowire region. The arrows mark the direction of increasing bismuth fraction as 3\%, 9\% and 15\%. The vertical dotted lines indicate the boundaries of the GaAs and \GaBiAs regions. The nanowire geometry parameters are $\rm D_S$ = 25 nm and $\rm D_C$ = 20 nm. \textbf{(b)} Same plots as (a) but for the nanowire geometry parameter as $\rm D_S$ = 40 nm and $\rm D_C$ = 20 nm. \textbf{(c)} The impact of the hydrostatic and biaxial strain components is schematically indicated on the lowest conduction band and the highest two valence band (heavy-hole and light-hole) edges is shown. The direction of the arrows indicates increase/decrease in the band edge energies due to increasing/decreasing magnitude in the strain component. The black(red) color of the arrows refers to hydrostatic(biaxial) component effect.  \textbf{(d, e)} Same plots as in (a) and (b) but for \GaAsGaBiAsGaAs nanowires with geometry parameters $\rm D_{OS}$ = 30 nm, $\rm D_{IS}$ = 25 nm and $\rm D_C$ = 20 nm in (d), and $\rm D_{OS}$ = 45 nm, $\rm D_{IS}$ = 40 nm and $\rm D_C$ = 20 nm in (e). \textbf{(f)} The same plot as in (c) but for a \GaAsGaBiAsGaAs nanowire.}
\label{fig:Fig2}
\end{figure*}  

In experimental studies, often multi$-$shell nanowires, also known as quantum well nanowires, are investigated where the nanowires are made up of two or more shell regions. In such nanowires, for example the outer shell and core regions can be made up of the same material composition whereas the inner shell designed with an alloy forming the active region of the nanowire for charge carrier confinements and/or transport. In this work, we also investigate the effect of GaAs outer shell forming \GaAsGaBiAsGaAs nanowires as shown in Figure~\ref{fig:Fig1} (e).  The band edge changes for such nanowires are schematically illustrated in Figure~\ref{fig:Fig1} (f), where the core and outer shell regions correspond to the larger band-gap GaAs material and the inner shell region is small band-gap \GaBiAs material which provides carrier confinement. The dependence of the band-gap energy (optical wavelength) on the Bi fraction inside the inner shell region is shown in Figure~\ref{fig:Fig1} (g), which exhibits a very similar trend as reported above for the \GaBiAsGaAs nanowires. We calculate approximately 29 meV and 36 meV reductions in the band-gap energies per \% Bi increase for $\rm D_{IS}$ = 25 nm and 40 nm, respectively. These changes are slightly smaller than computed for the \GaBiAsGaAs nanowires reported above. We also note that the reduction in the band-gap energy per \% increase in Bi composition for the investigated nanowires is significantly lower than the previously computed $\approx$90 meV reduction per \% Bi increase for the bulk \GaBiAs alloy. This can be attributed to the fact that for bulk \GaBiAs alloys, the incorporation of Bi strongly effect the band edge energies, in particular the valence band edge energy due to the band anti-crossing effect. However, for nanostrucures such as nanowires this band anti-crossing effect is modulated by other factors such as quantum confinement and internal strain which all contributes to the net reduction of the band-gap energy.

Figure~\ref{fig:Fig1} (h) plots the dependence of the band-gap energy (optical transition wavelength) for \GaAsGaBiAsGaAs nanowires as a function of nanowire diameter  $\rm D_{OS}$ by keeping the Bi fraction constant in the inner shell region. Interestingly, we compute a strong impact of the GaAs outer shell on the band-gap energy reduction when compared to the \GaBiAsGaAs nanowires. The band-gap energy decreases as a function of the $\rm D_{OS}$ diameter but the rate of decrease is noticeably slower. In particular, the band-gap energy becomes relatively flat at higher Bi fractions (12\% and 15\%). This implies that for \GaAsGaBiAsGaAs nanowires, the increase in the nanowire diameter $\rm D_{OS}$ will only negligibly shift the band-gap energy for Bi fractions above 12\%.  This non-trivial behaviour of nanowire band-gap reduction is a direct consequence of the internal strain modulation by the GaAs outer shell region, which will be discussed in the next strain analysis section.

We want to conclude this section by summarising that a large tunable band-gap energy (optical transition wavelength) is accessible from the investiagted \GaBiAsGaAs and \GaAsGaBiAsGaAs nanowires, which can be controlled by engineering geometry parameters (\textit{i.e.} nanowire diameters and/or Bi fractions). Importantly, we find that for Bi fractions around 15\%, the optical transition wavelength is in the spectral range of interest for telecommunication devices (1550 nm) which will be desirable for a variety of photonic devices. 
\noindent
\\ \\
\textbf{Strain Analysis}
\\ 
In order to understand the band-gap energy shifts discussed in the previous section, it is imperative to analyse strain profiles which directly affect the electron and hole energies and wave functions~\cite{Usman_IEEE_2009}. The nanowires studied in this work are made up of GaAs and \GaBiAs materials which have very different lattice constants -- 0.56532 nm for GaAs compared to 0.6328 nm for GaBi. The large lattice mismatch between these two materials gives rise to atomic bond deformations at the heterostructure interface, leading to an internal strain which penetrates deep in both GaAs core and \GaBiAs shell regions. The strain is responsible for shifting conduction and valence band edges~\cite{Usman_IEEE_2009}, leading to the variations in the band-gap energies. In this work, we have relaxed the atomistic structure of nanowires by using atomistic valence force field method~\cite{Keating_PR_1966, Usman_PRB_2011}, and the strain is computed from the relaxed atomic positions~\cite{Lee_1}. 

\begin{figure*}[t]
\includegraphics[scale=0.34]{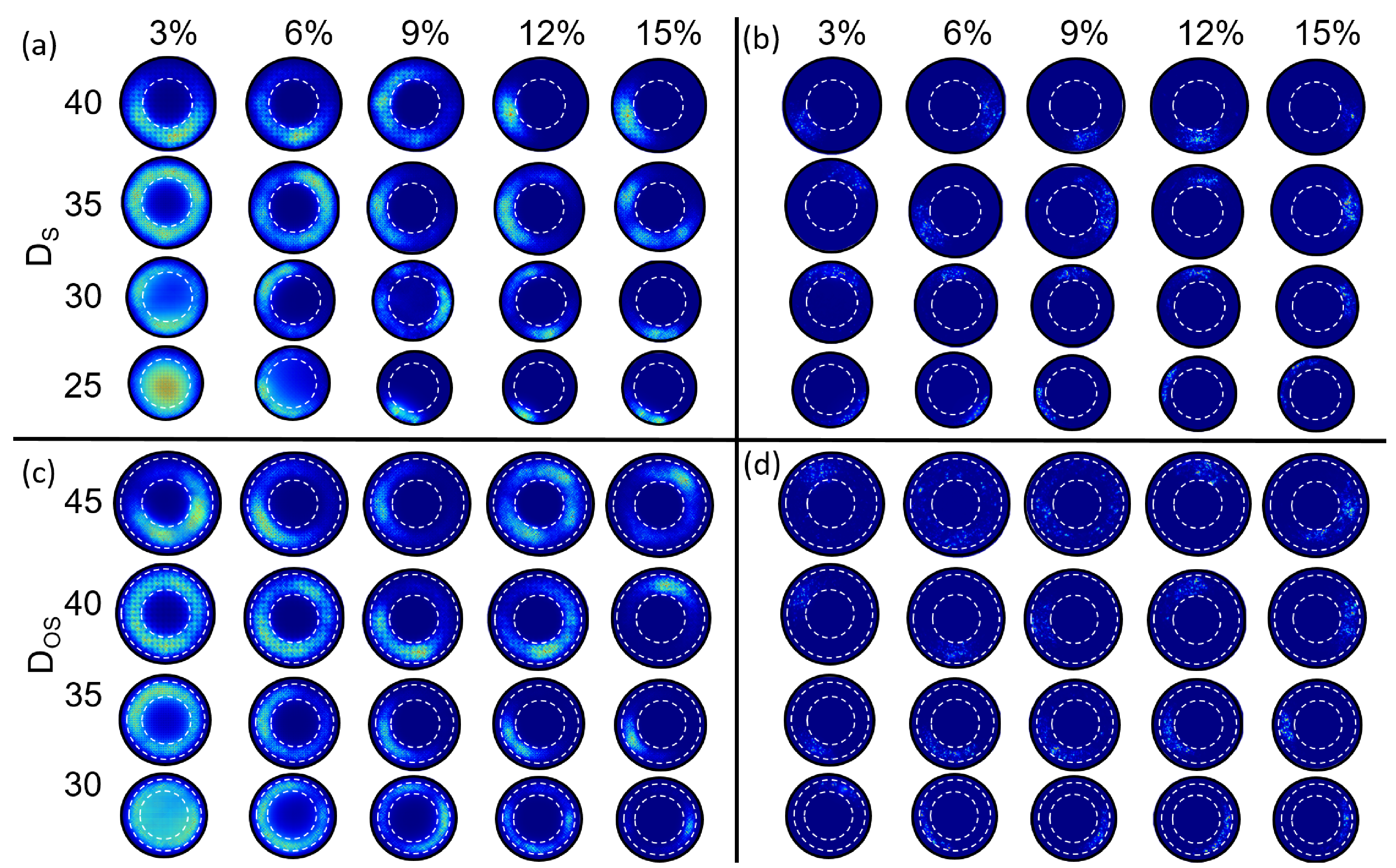}
\caption{\textbf{Wave function confinement.} \textbf{(a)} The spatial distribution of the lowest electron wave function confined in \GaBiAsGaAs nanowire is visualised in a (001) plane passing through the maxima. The blue color in the plots indicates low change density and the bright colors indicate high regions of charge density. The dotted circles are plotted to indicate the boundaries of the GaAs core regions. \textbf{(b)} Same as (a) but for the highest hole charge densities. \textbf{(c,d)} Same as (a) and (b) but for the \GaAsGaBiAsGaAs nanowires.}
\label{fig:Fig3}
\end{figure*}  

\begin{figure*}[t]
\includegraphics[scale=0.19]{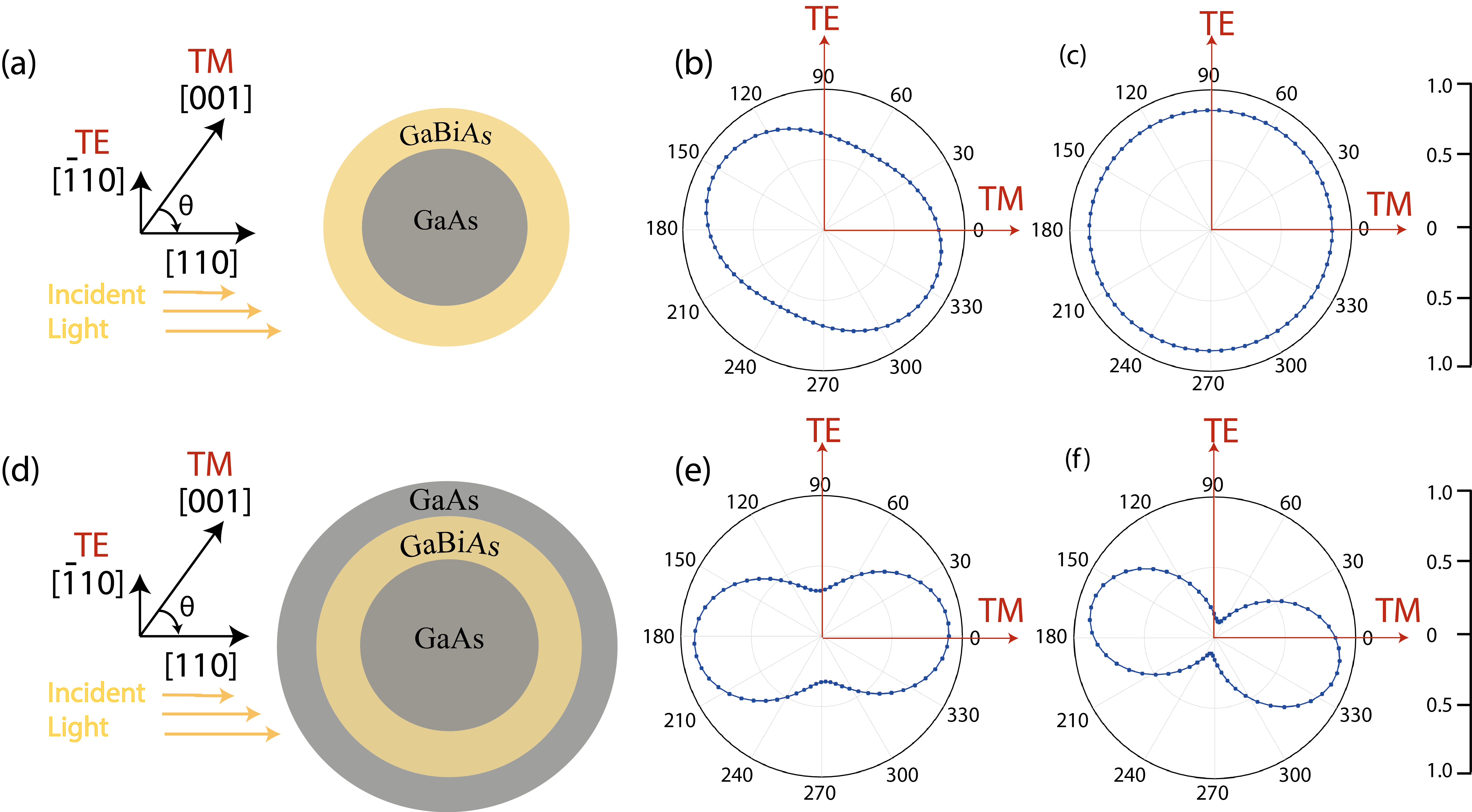}
\caption{\textbf{Polarisation-sensitive light absorption.} \textbf{(a)} The \GaBiAsGaAs nanowire geometry is illustrated in the [001] plane. To compute the polarisation-dependent optical transition strengths, the direction of the incident light is assumed to be along the [$\overline{1}$10] axis. The polarisation-dependent optical transition strengths are computed by varying the angle $\theta$ between the [001] and [110] directions. The inter-band transition strength at $\theta$=0(90$^o$) corresponds to TM(TE) mode. \textbf{(b)} The normalized polar plot is shown for the investigated nanowire with geometry parameters $\rm D_C$ = 20 nm, $\rm D_S$ = 40 nm and Bi=3\%. The red color arrows indicate the TE and TM mode intensities. \textbf{(c)} The same plot as in (b) but for the nanowire geometry parameters $\rm D_C$ = 20 nm, $\rm D_S$ = 40 nm, Bi=15\%. \textbf{(d)} The \GaAsGaBiAsGaAs nanowire geometery is illustrated in the [001] plane with the direction of incident light and polarisation angle $\theta$ marked as in (a). \textbf{(e)} The same plot as in (b) but for the nanowire with parameters $\rm D_C$ = 20 nm, $\rm D_{IS}$ = 40 nm, $\rm D_{OS}$ = 45 nm, and Bi=3\%. \textbf{(f)} The polar plot as in (c) but for the nanowire parameters $\rm D_C$ = 20 nm, $\rm D_{IS}$ = 40 nm, $\rm D_{OS}$ = 45 nm and Bi=15\%. In each case, the polar plots are based on the cumulative sum of the optical transition strengths between the lowest conduction band state and the highest three valence band states.}
\label{fig:Fig4}
\end{figure*}  

The impact of the strain on the conduction and valence band edges can be understood by plotting and analysing hydrostatic and biaxial strain components, where the hydrostatic strain component is computed as $\epsilon_H = \epsilon_{xx} + \epsilon_{yy} + \epsilon_{zz} $ and the biaxial strain component is computed from $\epsilon_B = \epsilon_{xx} + \epsilon_{yy} - 2\epsilon_{zz}$~\cite{Singh_JAP_1985}. We first investigate the hydrostatic and biaxial strain profiles in \GaBiAsGaAs nanowires by plotting strain graphs along the [100] direction through the center of the nanowires region. Figures~\ref{fig:Fig2} (a) and (b) shows the plots of the $\epsilon_H$ and $\epsilon_B$ profiles for two nanowire diameters, 25 nm and 40 nm, respectively. In each case, the strain profiles are plotted for three Bi fractions: 3\%, 9\% and 15\% as indicated by arrows. In general, we compute that the $\epsilon_H$ and $\epsilon_B$ strain profiles exhibit opposite signs both inside the core as well as inside the shell regions. This trend is consistent for all of the investigated nanowires and independent of the nanowire diameters and the Bi fractions. We also find the long-range penetration of the strain inside the core region which is heavily strained. 

Generally, the strain character at the interface of a small and large lattice constant materials is such that the larger lattice constant material is heavily compressed whereas the smaller lattice constant material is stretched. This results in compressive (negative) $\epsilon_H$ and $\epsilon_B$ within the larger lattice constant material and tensile (positive) $\epsilon_H$ and $\epsilon_B$ inside the smaller lattice constant material. This has been reported in many previous studies on quantum wells and quantum dots, for example GaAs/InAs quantum dots~\cite{Usman_IEEE_2009} and quantum dot molecules~\cite{Usman_JAP_2011_2}. From the strain profiles plotted in Figures~\ref{fig:Fig2} (a) and (b), we find that the calculated $\epsilon_H$ strain is tensile inside the GaAs core region and it is compressive in the \GaBiAs shell region. The character of the $\epsilon_B$ strain is opposite to that of the $\epsilon_H$ strain. We also note a couple of interesting properties from the plotted strain profiles: first, the $\epsilon_H$ strain is of equal magnitude to the $\epsilon_B$ strain when the nanowire diameter is 25 nm, which is not the case for nanowire diameter of 40 nm where the strain is relatively asymmetric and the biaxial strain is much smaller than the hydrostatic strain. Secondly, the magnitude of strain is much stronger in the \GaBiAs shell region when nanowire diameter is small, whereas at 40 nm diameter, the shell region is nearly strain free. This character of strain will have direct consequence for the confinement of electron and hole wave functions which will be discussed in the next section. The strain profiles also show that by varying the diameter of the nanowire in the growth process, the strain can be controlled which has a direct consequence for the electron hole energies and confinements. 

The impact of strain on the nanowire electronic structure is schematically shown in Figure~\ref{fig:Fig2} (c), where we have plotted the lowest conduction band edge and the highest two valence band edges, namely the heavy-hole (HH) and light-hole (LH) bands. The effect of the strain character (compressive or tensile) is shown on the corresponding band edges by indicating arrow directions, where the black arrows indicate the effect of $\epsilon_H$ strain and the red arrows show the impact of $\epsilon_B$ strain. The direction of the arrows indicates an increase or decrease in the corresponding band edge energy. Based on the strain profiles, the $\epsilon_H$ strain will generally increase the carrier confinement in the \GaBiAs shell region. The $\epsilon_B$ strain, on the other hand, only impacts the valence band states. It is important to note that for small diameter nanowires, the roughly equal magnitude of $\epsilon_B$ and $\epsilon_H$ implies that hole energies will experience a relatively small strain induced shift. On the other hand, for large diameter nanowires, $\epsilon_B$ is much smaller than $\epsilon_H$, and therefore the net impact of strain on hole energies is stronger. This is reflected in the band-gap energy shifts plotted in Figure~\ref{fig:Fig1} (d) where the red shift of the band-gap energy is relatively strong at larger nanowire diameters compared to the small diameters of nanowires. 

Another important effect of strain is related to light polarisation. Inside the \GaBiAs shell region where the carriers are confined, we compute that the presence of the tensile biaxial strain will enhance the mixing of LH character in the topmost valence band states. This is similar to large stacks of the quantum dots, where a very strong interdot coupling leads to a tensile strain which increases LH contribution of valence band states~\cite{Usman_PRB2_2011}. As we will show in the later section, this enhanced LH mixing in the valence states has strong impact on the polarisation sensitivity of the inter-band light absorption which is an important parameter for many optoelectronic devices.

In what follows, we investigate the impact of the GaAs outer shell by plotting $\epsilon_H$ and $\epsilon_B$ strain profiles for \GaAsGaBiAsGaAs nanowires in Figures~\ref{fig:Fig2} (d) and (e). To enable a direct comparison with \GaBiAsGaAs strain profiles, we have again selected the same Bi fractions and the same \GaBiAs shell diameters. Overall, we find that the strain magnitude is reduced when GaAs outer shell is present in the nanowire region. Other than that, the general trend and the tensile/compressive character of the $\epsilon_H$ and $\epsilon_B$ strain profiles remain same in different regions of the nanowires and therefore the insights gained above for \GaBiAsGaAs nanowires are largely applicable for multi$-$shell nanowires as well. The similar character of the $\epsilon_B$ and $\epsilon_H$ strain profiles mean their impact on the conduction and valence band edges is also same as schematically shown in Figure~\ref{fig:Fig2} (f). An important difference between the two nanowires is to note that the $\epsilon_B$ strain magnitude is relatively larger in \GaAsGaBiAsGaAs nanowires at 45 nm diameter when compared to \GaBiAsGaAs nanowires at 40 nm diameter. This implies that hole energies will experience a relatively small shift in energy for the multi$-$shell nanowires, which is responsible for the band-gap wavelength becomes flat as a function of the nanowire diameter in Figure~\ref{fig:Fig1} (h). Overall, we conclude that the presence of the GaAs outer shell region induces a small quantitative effect on the internal strain of the nanowires. 
\noindent
\\ \\
\textbf{Charge Carrier Confinements}
\\ 
The inter-band optical absorption strength depends on the spatial profiles of the confined electron and hole charge carriers in the nanowire region. Figure~\ref{fig:Fig3} (a) and (b) plots the spatial distribution of the lowest electron and the highest hole states through a [001] plane at the center of the \GaBiAsGaAs nanowire region. The dashed white circles indicate the boundary of the GaAs core region. Notably, our calculations show that at 3\% Bi composition and small nanowire diameters (25 nm), the electron wave function is confined inside the GaAs core region and the hole wave function is confined in the \GaBiAs shell region, leading to a quasi type-II confinement. This is attributed to a very small hydrostatic strain as shown in the Figure~\ref{fig:Fig2} (a), which leads to a negligibly small potential barrier for the electrons and therefore a significant leakage of the electron wave function in the GaAs region is computed. Overall, the calculations indicate that at 3\% Bi fraction, the electron wave function is distributed inside the large part of the nanowire region but becomes highly confined at larger Bi fractions. This can be understood from the strain profiles plotted in Figure~\ref{fig:Fig3} (a) and (b). For small nanowire diameters, the hydrostatic strain inside the GaAs core region is relatively small compared to \GaBiAs shell region. Therefore, the electron wave functions experience a small potential barrier at the GaAs and \GaBiAs interface and some wave function leakage is expected in the core region. However, when the nanowire diameter is increased, the magnitude of the hydrostatic strain drastically increases in the GaAs core region, leading to a very large potential barrier which prohibits any electron wave function leakage inside the core region. Therefore, the electron wave functions are much more confined inside the shell region. 

In comparison to the electron wave functions, the hole wave functions are impacted by both strain and alloy composition (pairs and clusters of Bi atoms)~\cite{Usman_APL_2014}. Previous studies on bismuth-containing nanowires~\cite{Usman_Nanoscale_2019} and quantum wells~\cite{Usman_PRM_2018, Usman_PRA_2018} have shown that the effect of alloy composition is stronger than the strain effect and therefore, the hole wave functions are heavily confined in the regions of Bi related clusters. In Figure~\ref{fig:Fig3} (b), we observe the same trend and the results show highly confined hole wave functions inside the \GaBiAs shell regions.

The effect of the GaAs outer shell region on charge carrier confinements is shown in Figures~\ref{fig:Fig3} (c) and (d), where we plot the lowest electron and the highest hole wave functions in the [001] plane through the center of the \GaAsGaBiAsGaAs nanowires for various diameters and Bi compositions. Overall, the effect of the GaAs outer shell is weak and the confinement of the electron and holes wave functions exhibit a very similar trend as noted earlier for the \GaBiAsGaAs nanowires. Again, for 3\% Bi fraction and 30 nm nanowire diameter, we compute a type-II confinement where the electron wave function is inside the GaAs core region and the hole wave function is confined inside the \GaBiAs shell region. As the diameter of nanowire is increased with fixed Bi composition, the electron wave function gradually becomes more confined inside the \GaBiAs shell region and the nanowire exhibits a type-I confinement. For the highest Bi composition (15\%), we find a strong electron wave function confinement inside the \GaBiAs shell region irrespective of nanowire diameter. The hole wave functions are again dominated by the Bi clustering effect and therefore exhibit a strong confinement inside the \GaBiAs shell region at all of the investigated diameters and compositions.
\\ \\
\noindent
\textbf{Polarisation-sensitive Light Absorption}
\\
The polarisation sensitivity of the inter-band optical transitions is an important parameter of interest for many photonic devices. In the past studies, this parameter has been investigated for many nanostructures, including quantum dots~\cite{Usman_PRB_2012}, quantum dot molecules~\cite{Usman_JAP_2011_2}, quantum dot stacks~\cite{Usman_PRB2_2011, Greif_ACSPhotonics_2018, Inoue_APL_2010}, nanowires~\cite{Wang_Science_2001, Casadei_SciRep_2015}, and nanorods~\cite{Hadar_JPCL_2013, Zhi_IEEEP_2016}. The polarisation response of a nanostructure is characterised in terms of polarisation ratio or degree of polarization (DOP) which is defined as $\frac{TE-TM}{TE+TM}$, where TE is transverse electric mode and TM is traverse magnetic mode of the incident polarised light.

Generally, for a lattice mismatched nanostructure where a large lattice constant material such as InAs is surrounded by a small lattice constant material such as GaAs, the strong compressive biaxial strain leads to high sensitivity of the light absorption towards TE mode polarisation~\cite{Usman_JAP_2011, Greif_ACSPhotonics_2018}, resulting in DOP values closer to 1. This is because the highly compressive biaxial strain shifts HH band towards higher energies and the LH band towards the lower energies, leading to the top-most valence band states dominated by HH character, which only couples to the TE mode. For polarisation-insensitive light interaction (\textit{i.e.} DOP=0), nanostructures are engineered to achieve tensile biaxial strain such that the LH mixing of the valence band increases TM mode coupling. This, for example, has been done in the past for large quantum dot stacks~\cite{Usman_PRB2_2011, Inoue_APL_2010, Greif_ACSPhotonics_2018}, where DOP values close to zero are reported both from experimental measurements and theoretical simulations. Here, in this work, we have shown that the biaxial strain is tensile inside the \GaBiAs shell region which naturally increases the LH character of the top-most hole states. As a result, a much larger TM mode coupling is expected for the investigated nanowires. 

Figure~\ref{fig:Fig4} plots the polarisation dependent optical transition strengths for a selected set of nanowires with 15\% Bi composition, which are relevant for telecommunication wavelength photonic devices. The incident light direction is assumed along the [110] direction, and the polarisation angle is varied between the [001] and [$\overline{1}$10] directions, where 0 angle indicates TM mode strength and 90 angle indicates TE mode strength. In all cases, we find a very strong TM mode strength which is a direct consequence of the LH mixing in the valence band states. Interestingly, we discover that for 15\% Bi composition, the \GaBiAsGaAs nanowire with 40 nm diameter exhibits a polarisation-insensitive light absorption (TE $\sim$ TM). This is very promising as the same nanowire has band-gap wavelength of 1550 nm which is directly relevant for the telecommunication wavelength devices. We also note that overall the polarisation response of \GaBiAsGaAs nanowires is more isotropic compared to the polarisation response of \GaAsGaBiAsGaAs nanowires, the latter being more dominated by TM mode. This is due to the biaxial strain profile studied earlier in Figures~\ref{fig:Fig2} (b) and (e), where the tensile biaxial strain is relatively stronger for the \GaAsGaBiAsGaAs nanowires which will result in higher mixing of LH band in the highest valence band states.
\\ \\
\noindent
\textbf{Summary and Outlook}
\\ \\
\noindent
\noindent
In summary, this work has investigated the optoelectronic properties of \GaBiAsGaAs and \GaAsGaBiAsGaAs core$-$shell nanowires by carrying out a systematic set of multi-million atom tight-binding calculations. The results show a highly tunable band-gap energy which can be engineered by either controlling the nanowire diameter or by adjusting the Bi composition in the shell region. Due to highly mismatched lattice parameters of GaAs and \GaBiAs materials, the strain is expected to have an impact comparable to quantum confinement, provided the interface remains relatively defect free and coherent. Remarkably, the detailed analysis of the internal strain profiles reveals a peculiar character which is found to profoundly impact the charge carrier confinements and sensitivity to light polarisation. The controllability of strain magnitude by Bi fractions and nanowire diameters allows modification of carrier confinements from a quasi type II to highly confined type I character, which broadens the scope of the investigated nanowires for myriad applications involving large carrier separations or strong inter-band optical transition strengths. A polarisation-dependent optical spectrum is computed which exhibits high TM mode coupling due to highly tensile strain character inside the \GaBiAs shell region. We also report the possibility of realising isotropic polarisation response at 1550 nm wavelength, which is highly desirable for many photonics devices such as optical amplifiers. Overall, our analysis uncovers important physical factors which impact the optoelectronic character of core$-$shell nanowires for investigated constituent materials and geometries, which can aid in the advancement of ongoing experimental efforts to incorporate \GaBiAs materials in the active region of nanowires.  

The idea of including \GaBiAs material in nanowires is relatively new, with the first experimental study published only recently~\cite{Ishikawa_Nanoletters_2015}. However, there is rapidly growing research interest in this novel material system as demonstrated by the subsequent experimental studies in the last few years showing promising progress~\cite{Steele_SciRep_2016, Zelewski_APL_2016, Wang_APE_2016, Oliva_arXiv_2019, Zhang_NanoLett_2019, Matsuda_NanoLett_2019, Matsuda_JAP_2019}. One of the critical challenges in the next few years will be to incorporate large Bi fractions in the active region of nanowire to achieve optical wavelengths to target photonic devices working in the near to mid and far infra-red regimes. For example as predicted in this work, the photonic devices aimed at the telecommunication applications would require Bi fractions in the range of 10 to 15\%. In the first experimental effort, about 2\% Bi was reported~\cite{Ishikawa_Nanoletters_2015} but recently upto 10\% Bi has been incorporated in the nanowire structure~\cite{Oliva_arXiv_2019}. We hope this study will further inspire and instigate experimental efforts to achieve larger Bi fractions. In short term with low Bi compositions ($\leq$ 3\%), these nanowires could also be explored for devices requiring quasi type II carrier confinements. We also note that there are other highly-mismatched bismuth-containing alloys such as GaSbBi, GaBiNAs, and InGaBiAs, which have been investigated in the literature and have shown very similar properties as observed for the GaBiAs alloy~\cite{Bismuth_containing_compounds_2013, Broderick_PSSB_2013, Rajpalke_APL_2013, Usman_PRA_2018, Bushell_SR_2019}. Although this work is focused only on the \GaBiAs material, it would be interesting to explore the optoelectronic properties of core$-$shell nanowires for a range of bismide materials in conjunction with experimental efforts. 
\\ \\
\noindent
\textbf{Methods}
\\ \\
\noindent
The electronic structure atomistic simulations performed in this work are based on a well-benchmarked tight-binding model~\cite{Usman_PRB_2011}, which have been verified against a number of experimental studies~\cite{}. The subsequent density functional theory calculations and experimental studies have also confirmed the accuracy of our tight-binding model. The tight-binding Hamiltonian is based on ten-band \textit{sp$^3$s$^*$} model including spin orbit coupling and is implemented within the framework of nanoelectronic modeling tool NEMO3D~\cite{Klimeck_IEEETED_2007_2}. The diagonalisation of the tight-binding Hamiltonian at the $\Gamma$ point provides electron and hole energies and wave functions which are plotted in Figures~\ref{fig:Fig1} and \ref{fig:Fig3}. Due to the lattice mismatch between GaAs and \GaBiAs materials, the atoms are displaced from their original lattice and lead to strain inside the nanostructure. The computation of strain is based on atomistic valence force field relaxation scheme~\cite{Keating_PR_1966}, where we have employed the experimentally reported bulk lattice constants and elastic coefficients for the GaAs and GaBi materials~\cite{Usman_PRB_2011}. The corresponding strain profiles are plotted in the Figure~\ref{fig:Fig2}. The inter-band optical transition strengths between the electron and hole wave functions are computed by using the Fermi's golden rule as follows: first we calculated squared absolute values of the momentum matrix elements summed over spin degenerate states, T$_{E_{i} \rightarrow H_{i}}$ = $|\langle E_{i}|[\overrightarrow{n},\mathbf{H}]|H_{i} \rangle |^{2}$ where $\mathbf{H}$ is the single particle tight binding Hamiltonian in the $sp^3s^*$ basis, $E_{i}$ and $H_{i}$ are the electron and hole energy states respectively, and $\overrightarrow{n}$ is a vector along the polarization direction. The polarization dependent polar plots are computed by rotating the polarization vector $\overrightarrow{n}$ = $(\overrightarrow{x}+\overrightarrow{y}) \cos \phi \sin \theta + \overrightarrow{z} \cos \theta $ to the appropriate direction in the polar coordinates: $\phi = 135^o$ and $\theta$ is varied from 0 to 360$^o$.

\begin{acknowledgement}

Computational resources are acknowledged from the National Computational Infrastructure (NCI) and Pawsey Supercomputer Center under the National Computational Merit based Allocation Scheme (NCMAS).

\end{acknowledgement}


\begin{mcitethebibliography}{74}
\providecommand*\natexlab[1]{#1}
\providecommand*\mciteSetBstSublistMode[1]{}
\providecommand*\mciteSetBstMaxWidthForm[2]{}
\providecommand*\mciteBstWouldAddEndPuncttrue
  {\def\EndOfBibitem{\unskip.}}
\providecommand*\mciteBstWouldAddEndPunctfalse
  {\let\EndOfBibitem\relax}
\providecommand*\mciteSetBstMidEndSepPunct[3]{}
\providecommand*\mciteSetBstSublistLabelBeginEnd[3]{}
\providecommand*\EndOfBibitem{}
\mciteSetBstSublistMode{f}
\mciteSetBstMaxWidthForm{subitem}{(\alph{mcitesubitemcount})}
\mciteSetBstSublistLabelBeginEnd
  {\mcitemaxwidthsubitemform\space}
  {\relax}
  {\relax}

\bibitem[Dasgupta et~al.(2014)Dasgupta, Sun, Liu, Brittman, Andrews, Lim, Gao,
  Yan, and Yang]{Dasgupta_advm_2014}
Dasgupta,~N.; Sun,~J.; Liu,~C.; Brittman,~S.; Andrews,~S.; Lim,~J.; Gao,~H.;
  Yan,~R.; Yang,~P. {25th Anniversary Article: Semiconductor Nanowires $–$
  Synthesis, Characterization, and Applications}. \emph{Adv. Mater.}
  \textbf{2014}, \emph{26}, 2137\relax
\mciteBstWouldAddEndPuncttrue
\mciteSetBstMidEndSepPunct{\mcitedefaultmidpunct}
{\mcitedefaultendpunct}{\mcitedefaultseppunct}\relax
\EndOfBibitem
\bibitem[Zhuang et~al.(2011)Zhuang, Ning, and Pan]{Zhuang_Advm_2011}
Zhuang,~X.; Ning,~C.; Pan,~A. {Composition and Bandgap$-$Graded Semiconductor
  Alloy Nanowires}. \emph{Adv. Mater.} \textbf{2011}, \emph{24}, 13\relax
\mciteBstWouldAddEndPuncttrue
\mciteSetBstMidEndSepPunct{\mcitedefaultmidpunct}
{\mcitedefaultendpunct}{\mcitedefaultseppunct}\relax
\EndOfBibitem
\bibitem[Yao et~al.(2013)Yao, Yan, and Lieber]{Yao_NNano_2013}
Yao,~J.; Yan,~H.; Lieber,~C. {A nanoscale combing technique for the large-scale
  assembly of highly aligned nanowires}. \emph{Nature Nanotechnology}
  \textbf{2013}, \emph{8}, 329\relax
\mciteBstWouldAddEndPuncttrue
\mciteSetBstMidEndSepPunct{\mcitedefaultmidpunct}
{\mcitedefaultendpunct}{\mcitedefaultseppunct}\relax
\EndOfBibitem
\bibitem[Tomioka et~al.(2012)Tomioka, Yoshimura, and
  Fukui]{Tomioka_Nature_2012}
Tomioka,~K.; Yoshimura,~M.; Fukui,~T. {A III$-$V nanowire channel on silicon
  for high-performance vertical transistors}. \emph{Nature} \textbf{2012},
  \emph{488}, 189\relax
\mciteBstWouldAddEndPuncttrue
\mciteSetBstMidEndSepPunct{\mcitedefaultmidpunct}
{\mcitedefaultendpunct}{\mcitedefaultseppunct}\relax
\EndOfBibitem
\bibitem[Chen et~al.(2011)Chen, Tran, Ng, Ko, Chuang, Sedgwick, and
  C.-Hasnain]{Chen_NatureP_2011}
Chen,~R.; Tran,~T.-T.; Ng,~K.; Ko,~W.; Chuang,~L.; Sedgwick,~F.; C.-Hasnain,~C.
  {Nanolasers grown on silicon}. \emph{Nature Photonics} \textbf{2011},
  \emph{5}, 170\relax
\mciteBstWouldAddEndPuncttrue
\mciteSetBstMidEndSepPunct{\mcitedefaultmidpunct}
{\mcitedefaultendpunct}{\mcitedefaultseppunct}\relax
\EndOfBibitem
\bibitem[Wang et~al.(2003)Wang, Song, Mu, Lieber, Whang, Jin, Wu, and
  Lieber]{Wang_Nanoletters_2003}
Wang,~D.; Song,~J.; Mu,~Y.; Lieber,~C.; Whang,~M.; Jin,~S.; Wu,~Y.; Lieber,~M.
  {Large-scale hierarchical organisation of nanowire arrays for integrated
  nanosystems}. \emph{Nano Lett.} \textbf{2003}, \emph{3}, 1255\relax
\mciteBstWouldAddEndPuncttrue
\mciteSetBstMidEndSepPunct{\mcitedefaultmidpunct}
{\mcitedefaultendpunct}{\mcitedefaultseppunct}\relax
\EndOfBibitem
\bibitem[Yan et~al.(2009)Yan, Gargas, and Yang]{Yan_Nature_Photo_2009}
Yan,~R.; Gargas,~D.; Yang,~P. {Nanowire Photonics}. \emph{Nature Photonics}
  \textbf{2009}, \emph{3}, 569\relax
\mciteBstWouldAddEndPuncttrue
\mciteSetBstMidEndSepPunct{\mcitedefaultmidpunct}
{\mcitedefaultendpunct}{\mcitedefaultseppunct}\relax
\EndOfBibitem
\bibitem[Eaton et~al.(2016)Eaton, Fu, Wong, Ning, and
  Yang]{Eaton_Nature_Review_2016}
Eaton,~S.; Fu,~A.; Wong,~A.; Ning,~C.-Z.; Yang,~P. {Semiconductor nanowire
  lasers}. \emph{Nature Reviews Materials} \textbf{2016}, \emph{1}, 16028\relax
\mciteBstWouldAddEndPuncttrue
\mciteSetBstMidEndSepPunct{\mcitedefaultmidpunct}
{\mcitedefaultendpunct}{\mcitedefaultseppunct}\relax
\EndOfBibitem
\bibitem[Cartoixa et~al.(2017)Cartoixa, Palummo, Hauge, Bakkers, and
  Rurali]{Cartoixa_Nanolett_2017}
Cartoixa,~X.; Palummo,~M.; Hauge,~H.~T.; Bakkers,~E. P. A.~M.; Rurali,~R.
  {Optical Emission in Hexagonal SiGe Nanowires}. \emph{Nano Lett.}
  \textbf{2017}, \emph{17}, 4753\relax
\mciteBstWouldAddEndPuncttrue
\mciteSetBstMidEndSepPunct{\mcitedefaultmidpunct}
{\mcitedefaultendpunct}{\mcitedefaultseppunct}\relax
\EndOfBibitem
\bibitem[Zhao et~al.(2020)Zhao, Nagashima, Zhang, Hosomi, Yoshida, Akihiro,
  Kanai, Mizukami, Zhu, Takahashi, Suzuki, Samransuksamer, Meng, Yasui, and
  Aoki]{Zhao_Nanolett_2020}
Zhao,~X.; Nagashima,~K.; Zhang,~G.; Hosomi,~T.; Yoshida,~H.; Akihiro,~Y.;
  Kanai,~M.; Mizukami,~W.; Zhu,~Z.; Takahashi,~T.; Suzuki,~M.;
  Samransuksamer,~B.; Meng,~G.; Yasui,~T.; Aoki,~Y. {Synthesis of
  Monodispersedly Sized ZnO Nanowires from Randomly Sized Seeds}. \emph{Nano
  Lett.} \textbf{2020}, \emph{20}, 599\relax
\mciteBstWouldAddEndPuncttrue
\mciteSetBstMidEndSepPunct{\mcitedefaultmidpunct}
{\mcitedefaultendpunct}{\mcitedefaultseppunct}\relax
\EndOfBibitem
\bibitem[Dimakis et~al.(2014)Dimakis, Jahn, Ramsteiner, Tahraoui, Grandal,
  Kong, Marquardt, Trampert, Riechert, and Geelhaar]{Dimakis_NanoLett_2014}
Dimakis,~E.; Jahn,~U.; Ramsteiner,~M.; Tahraoui,~A.; Grandal,~J.; Kong,~X.;
  Marquardt,~O.; Trampert,~A.; Riechert,~H.; Geelhaar,~L. {Coaxial Multishell
  {(In,Ga)As/GaAs} Nanowires for Near-Infrared Emission on Si Substrates}.
  \emph{Nano Lett.} \textbf{2014}, \emph{14}, 2604\relax
\mciteBstWouldAddEndPuncttrue
\mciteSetBstMidEndSepPunct{\mcitedefaultmidpunct}
{\mcitedefaultendpunct}{\mcitedefaultseppunct}\relax
\EndOfBibitem
\bibitem[Munshi et~al.(2012)Munshi, Dheeraj, Fauske, Kim, van Helvoort,
  Fimland, and Weman]{Munshi_NanoLett_2012}
Munshi,~A.; Dheeraj,~D.; Fauske,~V.; Kim,~D.-C.; van Helvoort,~A.;
  Fimland,~B.-O.; Weman,~H. {Vertically Aligned GaAs Nanowires on Graphite and
  Few-Layer Graphene: Generic Model and Epitaxial Growth}. \emph{Nano Lett.}
  \textbf{2012}, \emph{12}, 4570\relax
\mciteBstWouldAddEndPuncttrue
\mciteSetBstMidEndSepPunct{\mcitedefaultmidpunct}
{\mcitedefaultendpunct}{\mcitedefaultseppunct}\relax
\EndOfBibitem
\bibitem[Royo et~al.(2017)Royo, Luca, Rurali, and Zardo]{Royo_JPDAP_2017}
Royo,~M.; Luca,~M.; Rurali,~R.; Zardo,~I. {A review on {III$–$V}
  core$–$multishell nanowires: growth, properties, and applications}.
  \emph{J. Phys. D: Appl. Phys.} \textbf{2017}, \emph{50}, 143001\relax
\mciteBstWouldAddEndPuncttrue
\mciteSetBstMidEndSepPunct{\mcitedefaultmidpunct}
{\mcitedefaultendpunct}{\mcitedefaultseppunct}\relax
\EndOfBibitem
\bibitem[Balagula et~al.(2020)Balagula, Jansson, Yukimune, Stehr, Ishikawa,
  Chen, and Buyanova]{Balagula_SciRep_2020}
Balagula,~R.; Jansson,~M.; Yukimune,~M.; Stehr,~J.; Ishikawa,~F.; Chen,~W.;
  Buyanova,~I. {Effects of thermal annealing on localization and strain in
  core/ multishell GaAs/GaNAs/GaAs nanowires}. \emph{Sci. Rep.} \textbf{2020},
  \emph{10}, 8216\relax
\mciteBstWouldAddEndPuncttrue
\mciteSetBstMidEndSepPunct{\mcitedefaultmidpunct}
{\mcitedefaultendpunct}{\mcitedefaultseppunct}\relax
\EndOfBibitem
\bibitem[Kim et~al.(2006)Kim, Joyce, Gao, Tan, Jagadish, Paladugu, Zou, and
  Suvorova]{Kim_Nanoletters_2006}
Kim,~Y.; Joyce,~H.; Gao,~Q.; Tan,~H.; Jagadish,~C.; Paladugu,~M.; Zou,~J.;
  Suvorova,~A. {Influence of Nanowire Density on the Shape and Optical
  Properties of Ternary InGaAs Nanowires}. \emph{Nano Lett.} \textbf{2006},
  \emph{6}, 599\relax
\mciteBstWouldAddEndPuncttrue
\mciteSetBstMidEndSepPunct{\mcitedefaultmidpunct}
{\mcitedefaultendpunct}{\mcitedefaultseppunct}\relax
\EndOfBibitem
\bibitem[Ren et~al.(2014)Ren, Hu, Zhang, Zhu, Zhuang, Ma, Fan, Liao, Duan, and
  Pan]{Ren_Adv_Mat_2014}
Ren,~P.; Hu,~W.; Zhang,~Q.; Zhu,~X.; Zhuang,~X.; Ma,~L.; Fan,~X.; Liao,~H.;
  Duan,~X.; Pan,~A. {Band$-$Selective Infrared Photodetectors with
  Complete$-$Composition$-$Range InAs$_x$P$_{1‐x}$ Alloy Nanowires}.
  \emph{Adv. Mater.} \textbf{2014}, \emph{26}, 7444\relax
\mciteBstWouldAddEndPuncttrue
\mciteSetBstMidEndSepPunct{\mcitedefaultmidpunct}
{\mcitedefaultendpunct}{\mcitedefaultseppunct}\relax
\EndOfBibitem
\bibitem[Ma et~al.(2014)Ma, Hu, Zhang, Ren, Zhuang, Zhou, Xu, Li, Shan, Wang,
  Liao, Xu, and Pan]{Ma_Nano_Lett_2014}
Ma,~L.; Hu,~W.; Zhang,~Q.; Ren,~P.; Zhuang,~X.; Zhou,~H.; Xu,~J.; Li,~H.;
  Shan,~Z.; Wang,~X.; Liao,~L.; Xu,~H.; Pan,~A. {Room-Temperature Near-Infrared
  Photodetectors Based on Single Heterojunction Nanowires}. \emph{Nano Lett.}
  \textbf{2014}, \emph{14}, 694\relax
\mciteBstWouldAddEndPuncttrue
\mciteSetBstMidEndSepPunct{\mcitedefaultmidpunct}
{\mcitedefaultendpunct}{\mcitedefaultseppunct}\relax
\EndOfBibitem
\bibitem[Dai et~al.(2014)Dai, Zhang, Wang, Adamo, Liu, Huang, Couteau, and
  Soci]{Dai_NanoLett_2014}
Dai,~X.; Zhang,~S.; Wang,~Z.; Adamo,~G.; Liu,~H.; Huang,~Y.; Couteau,~C.;
  Soci,~C. {GaAs/AlGaAs Nanowire Photodetector}. \emph{Nano Lett.}
  \textbf{2014}, \emph{14}, 2688\relax
\mciteBstWouldAddEndPuncttrue
\mciteSetBstMidEndSepPunct{\mcitedefaultmidpunct}
{\mcitedefaultendpunct}{\mcitedefaultseppunct}\relax
\EndOfBibitem
\bibitem[Ishikawa et~al.(2015)Ishikawa, Akamatsu, Watanabe, Uesugi, Asahina,
  Jahn, and Shimomura]{Ishikawa_Nanoletters_2015}
Ishikawa,~F.; Akamatsu,~Y.; Watanabe,~K.; Uesugi,~F.; Asahina,~S.; Jahn,~U.;
  Shimomura,~S. {Metamorphic GaAs/GaAsBi Heterostructured Nanowires}.
  \emph{Nano Letters} \textbf{2015}, \emph{15}, 7265\relax
\mciteBstWouldAddEndPuncttrue
\mciteSetBstMidEndSepPunct{\mcitedefaultmidpunct}
{\mcitedefaultendpunct}{\mcitedefaultseppunct}\relax
\EndOfBibitem
\bibitem[Steele et~al.(2016)Steele, Lewis, Horvat, Nancarrow, Henini, Fan,
  Mazur, Schmidbauer, Ware, Yu, and Salamo]{Steele_SciRep_2016}
Steele,~J.~A.; Lewis,~R.~A.; Horvat,~J.; Nancarrow,~M. J.~B.; Henini,~M.;
  Fan,~D.; Mazur,~Y.~I.; Schmidbauer,~M.; Ware,~M.~E.; Yu,~S.-Q.; Salamo,~G.~J.
  {Surface effects of vapour-liquid-solid driven Bi surface droplets formed
  during molecular-beam-epitaxy of GaAsBi}. \emph{Sci. Rep.} \textbf{2016},
  \emph{6}, 28860\relax
\mciteBstWouldAddEndPuncttrue
\mciteSetBstMidEndSepPunct{\mcitedefaultmidpunct}
{\mcitedefaultendpunct}{\mcitedefaultseppunct}\relax
\EndOfBibitem
\bibitem[Zelewski et~al.(2016)Zelewski, Kopaczek, Linhart, Ishikawa, Shimomura,
  and Kudrawiec]{Zelewski_APL_2016}
Zelewski,~S.; Kopaczek,~J.; Linhart,~W.~M.; Ishikawa,~F.; Shimomura,~S.;
  Kudrawiec,~R. {Photoacoustic spectroscopy of absorption edge for GaAsBi/GaAs
  nanowires grown on Si substrate}. \emph{Appl. Phys. Lett.} \textbf{2016},
  \emph{109}, 182106\relax
\mciteBstWouldAddEndPuncttrue
\mciteSetBstMidEndSepPunct{\mcitedefaultmidpunct}
{\mcitedefaultendpunct}{\mcitedefaultseppunct}\relax
\EndOfBibitem
\bibitem[Wang et~al.(2016)Wang, Pan, Wu, Cao, Wang, and Gong]{Wang_APE_2016}
Wang,~P.; Pan,~W.; Wu,~X.; Cao,~C.; Wang,~S.; Gong,~Q. {Heteroepitaxy growth of
  GaAsBi on Ge(100) substrate by gas source molecular beam epitaxy}.
  \emph{Appl. Phys. Express} \textbf{2016}, \emph{9}, 045502\relax
\mciteBstWouldAddEndPuncttrue
\mciteSetBstMidEndSepPunct{\mcitedefaultmidpunct}
{\mcitedefaultendpunct}{\mcitedefaultseppunct}\relax
\EndOfBibitem
\bibitem[Oliva et~al.(2019)Oliva, Gao, Luna, Geelhaar, and
  Lewis]{Oliva_arXiv_2019}
Oliva,~M.; Gao,~G.; Luna,~E.; Geelhaar,~L.; Lewis,~R. {Axial GaAs/Ga(As,Bi)
  Nanowire Heterostructures}. \emph{Nanotechnology} \textbf{2019}, \emph{30},
  425601\relax
\mciteBstWouldAddEndPuncttrue
\mciteSetBstMidEndSepPunct{\mcitedefaultmidpunct}
{\mcitedefaultendpunct}{\mcitedefaultseppunct}\relax
\EndOfBibitem
\bibitem[Zhang et~al.(2019)Zhang, Huang, Stehr, Chen, Wang, Lu, Chen, and
  Buyanova]{Zhang_NanoLett_2019}
Zhang,~B.; Huang,~Y.; Stehr,~J.~E.; Chen,~P.-P.; Wang,~X.-J.; Lu,~W.;
  Chen,~W.~M.; Buyanova,~I.~A. {Band Structure of Wurtzite GaBiAs Nanowires}.
  \emph{Nano Lett.} \textbf{2019}, \emph{19}, 6454\relax
\mciteBstWouldAddEndPuncttrue
\mciteSetBstMidEndSepPunct{\mcitedefaultmidpunct}
{\mcitedefaultendpunct}{\mcitedefaultseppunct}\relax
\EndOfBibitem
\bibitem[Matsuda et~al.(2019)Matsuda, Takada, Yano, Tsutsumi, Yoshikawa,
  Shimomura, Shimizu, Nagashima, Yanagida, and Ishikawa]{Matsuda_NanoLett_2019}
Matsuda,~T.; Takada,~K.; Yano,~K.; Tsutsumi,~R.; Yoshikawa,~K.; Shimomura,~S.;
  Shimizu,~Y.; Nagashima,~K.; Yanagida,~T.; Ishikawa,~F. {Controlling
  Bi-Provoked Nanostructure Formation in GaAs/GaAsBi Core-Shell Nanowires}.
  \emph{Nano Lett.} \textbf{2019}, \emph{19}, 8510\relax
\mciteBstWouldAddEndPuncttrue
\mciteSetBstMidEndSepPunct{\mcitedefaultmidpunct}
{\mcitedefaultendpunct}{\mcitedefaultseppunct}\relax
\EndOfBibitem
\bibitem[Matsuda et~al.(2019)Matsuda, Takada, Yano, Shimomura, and
  Ishikawa]{Matsuda_JAP_2019}
Matsuda,~T.; Takada,~K.; Yano,~K.; Shimomura,~S.; Ishikawa,~F. {Strain
  deformation in GaAs/GaAsBi core-shell nanowire heterostructures}. \emph{J.
  Appl. Phys.} \textbf{2019}, \emph{125}, 194301\relax
\mciteBstWouldAddEndPuncttrue
\mciteSetBstMidEndSepPunct{\mcitedefaultmidpunct}
{\mcitedefaultendpunct}{\mcitedefaultseppunct}\relax
\EndOfBibitem
\bibitem[Janotti et~al.(2002)Janotti, Wei, and Zhang]{Janotti_PRB_2002}
Janotti,~A.; Wei,~S.-H.; Zhang,~S.~B. Theoretical study of the effects of
  isovalent coalloying of {Bi} and {N} in {GaAs}. \emph{Phys.~Rev.~B}
  \textbf{2002}, \emph{65}, 115203\relax
\mciteBstWouldAddEndPuncttrue
\mciteSetBstMidEndSepPunct{\mcitedefaultmidpunct}
{\mcitedefaultendpunct}{\mcitedefaultseppunct}\relax
\EndOfBibitem
\bibitem[Zhang et~al.(2005)Zhang, Mascarenhas, and Wang]{Zhang_PRB_2005}
Zhang,~Y.; Mascarenhas,~A.; Wang,~L.-W. Similar and dissimilar aspects of
  {III$-$V} semiconductors containing Bi versus N. \emph{Phys. Rev. B}
  \textbf{2005}, \emph{71}, 155201\relax
\mciteBstWouldAddEndPuncttrue
\mciteSetBstMidEndSepPunct{\mcitedefaultmidpunct}
{\mcitedefaultendpunct}{\mcitedefaultseppunct}\relax
\EndOfBibitem
\bibitem[Usman et~al.(2011)Usman, Broderick, Lindsay, and
  O'Reilly]{Usman_PRB_2011}
Usman,~M.; Broderick,~C.~A.; Lindsay,~A.; O'Reilly,~E.~P. Tight-binding
  analysis of the electronic structure of dilute bismide alloys of {GaP} and
  {GaAs}. \emph{Phys.~Rev.~B} \textbf{2011}, \emph{84}, 245202\relax
\mciteBstWouldAddEndPuncttrue
\mciteSetBstMidEndSepPunct{\mcitedefaultmidpunct}
{\mcitedefaultendpunct}{\mcitedefaultseppunct}\relax
\EndOfBibitem
\bibitem[Li and Wang(2013)Li, and Wang]{Bismuth_containing_compounds_2013}
Li,~H., Wang,~Z., Eds. \emph{Bismuth-Containing Compounds}; Springer-Verlag New
  York, 2013\relax
\mciteBstWouldAddEndPuncttrue
\mciteSetBstMidEndSepPunct{\mcitedefaultmidpunct}
{\mcitedefaultendpunct}{\mcitedefaultseppunct}\relax
\EndOfBibitem
\bibitem[Broderick et~al.(2017)Broderick, Marko, O'Reilly, and
  Sweeney]{Broderick_bismide_chapter_2017}
Broderick,~C.~A.; Marko,~I.~P.; O'Reilly,~E.~P.; Sweeney,~S.~J. \emph{``Dilute
  Bismide Alloys'', Chapter 10 -- Handbook of Optoelectronic Device Modeling
  and Simulation (Vol.~1), in press} \textbf{2017}, \relax
\mciteBstWouldAddEndPunctfalse
\mciteSetBstMidEndSepPunct{\mcitedefaultmidpunct}
{}{\mcitedefaultseppunct}\relax
\EndOfBibitem
\bibitem[Fluegel et~al.(2006)Fluegel, Francoeur, Mascarenhas, Tixier, Young,
  and Tiedje]{Fluegel_PRL_2006}
Fluegel,~B.; Francoeur,~S.; Mascarenhas,~A.; Tixier,~S.; Young,~E.; Tiedje,~T.
  {Giant Spin-Orbit Bowing in GaAs$_{1-x}$Bi$_x$}. \emph{Phys. Rev. Lett.}
  \textbf{2006}, \emph{97}, 067205\relax
\mciteBstWouldAddEndPuncttrue
\mciteSetBstMidEndSepPunct{\mcitedefaultmidpunct}
{\mcitedefaultendpunct}{\mcitedefaultseppunct}\relax
\EndOfBibitem
\bibitem[Tumenas et~al.(2012)Tumenas, Karpus, Bertulis, and
  Arwin]{Tumenas_PSSc_2012}
Tumenas,~S.; Karpus,~V.; Bertulis,~K.; Arwin,~H. {Dielectric function and
  refractive index of GaBi$_x$As$_{1‐x}$ (x = 0.035, 0.052, 0.075)}.
  \emph{Phys. Status Solidi c} \textbf{2012}, \emph{9}, 1633\relax
\mciteBstWouldAddEndPuncttrue
\mciteSetBstMidEndSepPunct{\mcitedefaultmidpunct}
{\mcitedefaultendpunct}{\mcitedefaultseppunct}\relax
\EndOfBibitem
\bibitem[Batool et~al.(2012)Batool, Hild, Hosea, Lu, Tiedje, and
  Sweeney]{Batool_JAP_2012}
Batool,~Z.; Hild,~K.; Hosea,~T. J.~C.; Lu,~X.; Tiedje,~T.; Sweeney,~S.~J. The
  electronic band structure of {GaBiAs/GaAs} layers: {I}nfluence of strain and
  band anti-crossing. \emph{J. Appl. Phys.} \textbf{2012}, \emph{111},
  113108\relax
\mciteBstWouldAddEndPuncttrue
\mciteSetBstMidEndSepPunct{\mcitedefaultmidpunct}
{\mcitedefaultendpunct}{\mcitedefaultseppunct}\relax
\EndOfBibitem
\bibitem[Phillips et~al.(1999)Phillips, Sweeney, Adams, and
  Thijs]{Phillips_IEEEJSTQE_1999}
Phillips,~A.~F.; Sweeney,~S.~J.; Adams,~A.~R.; Thijs,~P.~J.~A. The temperature
  dependence of 1.3- and 1.5-$\mu$m compressively strained {InGaAs(P)} {MQW}
  semiconductor lasers. \emph{IEEE J.~Sel.~Top.~Quantum Electron.}
  \textbf{1999}, \emph{5}, 401\relax
\mciteBstWouldAddEndPuncttrue
\mciteSetBstMidEndSepPunct{\mcitedefaultmidpunct}
{\mcitedefaultendpunct}{\mcitedefaultseppunct}\relax
\EndOfBibitem
\bibitem[Sweeney et~al.(1999)Sweeney, Adams, Silver, O'Reilly, Watling, Walker,
  and Thijs]{Sweeney_PSSB_1999}
Sweeney,~S.~J.; Adams,~A.~R.; Silver,~M.; O'Reilly,~E.~P.; Watling,~J.~R.;
  Walker,~A.~B.; Thijs,~P.~J.~A. Dependence of threshold current on {QW}
  position and on pressure in 1.5 $\mu$m {InGaAs(P)} lasers. \emph{Phys.~Status
  Solidi B} \textbf{1999}, \emph{211}, 525\relax
\mciteBstWouldAddEndPuncttrue
\mciteSetBstMidEndSepPunct{\mcitedefaultmidpunct}
{\mcitedefaultendpunct}{\mcitedefaultseppunct}\relax
\EndOfBibitem
\bibitem[Broderick et~al.(2012)Broderick, Usman, Sweeney, and
  O'Reilly]{Broderick_SST_2012}
Broderick,~C.~A.; Usman,~M.; Sweeney,~S.~J.; O'Reilly,~E.~P. Band engineering
  in dilute nitride and bismide semiconductor lasers. \emph{Semicond. Sci.
  Technol.} \textbf{2012}, \emph{27}, 094011\relax
\mciteBstWouldAddEndPuncttrue
\mciteSetBstMidEndSepPunct{\mcitedefaultmidpunct}
{\mcitedefaultendpunct}{\mcitedefaultseppunct}\relax
\EndOfBibitem
\bibitem[Usman(2019)]{Usman_Nanoscale_2019}
Usman,~M. {Towards low-loss telecom-wavelength photonic devices by designing
  GaBiAs/GaAs core-shell nanowires}. \emph{Nanoscale} \textbf{2019}, \emph{11},
  20133\relax
\mciteBstWouldAddEndPuncttrue
\mciteSetBstMidEndSepPunct{\mcitedefaultmidpunct}
{\mcitedefaultendpunct}{\mcitedefaultseppunct}\relax
\EndOfBibitem
\bibitem[Usman et~al.(2013)Usman, Broderick, Batool, Hild, Hosea, Sweeney, and
  O'Reilly]{Usman_PRB_2013}
Usman,~M.; Broderick,~C.~A.; Batool,~Z.; Hild,~K.; Hosea,~T. J.~C.;
  Sweeney,~S.~J.; O'Reilly,~E.~P. Impact of alloy disorder on the band
  structure of compressively strained {GaBi$_{x}$As$_{1-x}$}. \emph{Phys. Rev.
  B} \textbf{2013}, \emph{87}, 115104\relax
\mciteBstWouldAddEndPuncttrue
\mciteSetBstMidEndSepPunct{\mcitedefaultmidpunct}
{\mcitedefaultendpunct}{\mcitedefaultseppunct}\relax
\EndOfBibitem
\bibitem[Usman et~al.(2018)Usman, Broderick, and O'Reilly]{Usman_PRA_2018}
Usman,~M.; Broderick,~C.~A.; O'Reilly,~E.~P. {Impact of disorder on the
  optoelectronic properties of {GaN$_y$As$_{1-x-y}$Bi$_x$} alloys and
  heterostructures}. \emph{Phys. Rev. Applied} \textbf{2018}, \emph{10},
  044024\relax
\mciteBstWouldAddEndPuncttrue
\mciteSetBstMidEndSepPunct{\mcitedefaultmidpunct}
{\mcitedefaultendpunct}{\mcitedefaultseppunct}\relax
\EndOfBibitem
\bibitem[Kudrawiec et~al.(2014)Kudrawiec, Kopaczek, Polak, Scharoch,
  Gladysiewicz, Misiewicz, Richards, Bastiman, and David]{Kudrawiec_JAP_2014}
Kudrawiec,~R.; Kopaczek,~J.; Polak,~M.; Scharoch,~P.; Gladysiewicz,~M.;
  Misiewicz,~J.; Richards,~R.; Bastiman,~F.; David,~J. \emph{J. Appl. Phys.}
  \textbf{2014}, \emph{116}, 233508\relax
\mciteBstWouldAddEndPuncttrue
\mciteSetBstMidEndSepPunct{\mcitedefaultmidpunct}
{\mcitedefaultendpunct}{\mcitedefaultseppunct}\relax
\EndOfBibitem
\bibitem[Polak et~al.(2015)Polak, Scharoch, and Kudrawiec]{Polak_SST_2015}
Polak,~M.~P.; Scharoch,~P.; Kudrawiec,~R. \emph{Semicond. Sci. Technol.}
  \textbf{2015}, \emph{30}, 094001\relax
\mciteBstWouldAddEndPuncttrue
\mciteSetBstMidEndSepPunct{\mcitedefaultmidpunct}
{\mcitedefaultendpunct}{\mcitedefaultseppunct}\relax
\EndOfBibitem
\bibitem[Bannow et~al.(2016)Bannow, Rubel, Badescu, Rosenow, Hader, Moloney,
  Tonner, and Koch]{Bannow_PRB_2016}
Bannow,~L.~C.; Rubel,~O.; Badescu,~S.; Rosenow,~P.; Hader,~J.; Moloney,~J.;
  Tonner,~R.; Koch,~S. \emph{Phys. Rev. B} \textbf{2016}, \emph{93},
  205202\relax
\mciteBstWouldAddEndPuncttrue
\mciteSetBstMidEndSepPunct{\mcitedefaultmidpunct}
{\mcitedefaultendpunct}{\mcitedefaultseppunct}\relax
\EndOfBibitem
\bibitem[Donmez et~al.(2015)Donmez, Erol, Arikan, Makhloufi, Arnoult, and
  Fontaine]{Donmez_SST_2015}
Donmez,~O.; Erol,~A.; Arikan,~M.; Makhloufi,~H.; Arnoult,~A.; Fontaine,~C.
  \emph{Semicond. Sci. Technol.} \textbf{2015}, \emph{30}, 094016\relax
\mciteBstWouldAddEndPuncttrue
\mciteSetBstMidEndSepPunct{\mcitedefaultmidpunct}
{\mcitedefaultendpunct}{\mcitedefaultseppunct}\relax
\EndOfBibitem
\bibitem[Balanta et~al.(2017)Balanta, Gordo, Carvalho, Puustinen, Alghamdi,
  Henini, Galeti, Guina, and Gobato]{Balanta_JoL_2017}
Balanta,~M. A.~G.; Gordo,~V.~O.; Carvalho,~A. R.~H.; Puustinen,~J.;
  Alghamdi,~H.~M.; Henini,~M.; Galeti,~H. V.~A.; Guina,~M.; Gobato,~Y.~G.
  {Polarization resolved photoluminescence in GaAs1-xBix/GaAs quantum wells}.
  \emph{J. of Luminescence} \textbf{2017}, \emph{182}, 49--52\relax
\mciteBstWouldAddEndPuncttrue
\mciteSetBstMidEndSepPunct{\mcitedefaultmidpunct}
{\mcitedefaultendpunct}{\mcitedefaultseppunct}\relax
\EndOfBibitem
\bibitem[Zhang et~al.(2018)Zhang, Qiu, Chen, and Wang]{Zhang_JAP_2018}
Zhang,~B.; Qiu,~W.-Y.; Chen,~P.-P.; Wang,~X.-J. \emph{J. Appl. Phys.}
  \textbf{2018}, \emph{123}, 035702\relax
\mciteBstWouldAddEndPuncttrue
\mciteSetBstMidEndSepPunct{\mcitedefaultmidpunct}
{\mcitedefaultendpunct}{\mcitedefaultseppunct}\relax
\EndOfBibitem
\bibitem[Dybala et~al.(2017)Dybala, Kopaczek, Gladysiewicz, Pavelescu,
  Romanitan, Ligor, Arnoult, Fontaine, and Kudrawiec]{Dybala_APL_2017}
Dybala,~F.; Kopaczek,~J.; Gladysiewicz,~M.; Pavelescu,~E.-M.; Romanitan,~C.;
  Ligor,~O.; Arnoult,~A.; Fontaine,~C.; Kudrawiec,~R. \emph{J. of Appl. Phys.}
  \textbf{2017}, \emph{111}, 192104\relax
\mciteBstWouldAddEndPuncttrue
\mciteSetBstMidEndSepPunct{\mcitedefaultmidpunct}
{\mcitedefaultendpunct}{\mcitedefaultseppunct}\relax
\EndOfBibitem
\bibitem[Collar et~al.(2017)Collar, Li, Jiao, Guan, Losurdo, Humlicek, and
  Brown]{Collar_AIPA_2017}
Collar,~K.; Li,~J.; Jiao,~W.; Guan,~Y.; Losurdo,~M.; Humlicek,~J.; Brown,~A.~S.
  \emph{AIP Advances} \textbf{2017}, \emph{7}, 075016\relax
\mciteBstWouldAddEndPuncttrue
\mciteSetBstMidEndSepPunct{\mcitedefaultmidpunct}
{\mcitedefaultendpunct}{\mcitedefaultseppunct}\relax
\EndOfBibitem
\bibitem[Broderick et~al.(2014)Broderick, Mazzucato, Carr\`{e}re, Amand,
  Makhloufi, Arnoult, Fontaine, Donmez, Erol, Usman, O'Reilly, and
  Marie]{Broderick_PRB_2014}
Broderick,~C.~A.; Mazzucato,~S.; Carr\`{e}re,~H.; Amand,~T.; Makhloufi,~H.;
  Arnoult,~A.; Fontaine,~C.; Donmez,~O.; Erol,~A.; Usman,~M.; O'Reilly,~E.~P.;
  Marie,~X. Anisotropic electron g factor as a probe of the electronic
  structure of {GaBi$_{x}$As$_{1-x}$/GaAs} epilayers. \emph{Phys.~Rev.~B}
  \textbf{2014}, \emph{90}, 195301\relax
\mciteBstWouldAddEndPuncttrue
\mciteSetBstMidEndSepPunct{\mcitedefaultmidpunct}
{\mcitedefaultendpunct}{\mcitedefaultseppunct}\relax
\EndOfBibitem
\bibitem[Broderick et~al.(2013)Broderick, Usman, and
  O'Reilly]{Broderick_SST_2013}
Broderick,~C.~A.; Usman,~M.; O'Reilly,~E.~P. Derivation of 12 and 14-band
  \textbf{k}$\cdot$\textbf{p} Hamiltonians for dilute bismide and
  bismide-nitride alloys. \emph{Semicond.~Sci.~Technol.} \textbf{2013},
  \emph{28}, 125025\relax
\mciteBstWouldAddEndPuncttrue
\mciteSetBstMidEndSepPunct{\mcitedefaultmidpunct}
{\mcitedefaultendpunct}{\mcitedefaultseppunct}\relax
\EndOfBibitem
\bibitem[Broderick et~al.(2015)Broderick, Harnedy, Ludewig, Bushell, Volz,
  Manning, and O'Reilly]{Broderick_SST_2015}
Broderick,~C.~A.; Harnedy,~P.~E.; Ludewig,~P.; Bushell,~Z.~L.; Volz,~K.;
  Manning,~R.~J.; O'Reilly,~E.~P. Determination of type-{I} band offsets in
  {GaBi$_{x}$As$_{1-x}$} quantum wells using polarisation-resolved photovoltage
  measurements and 12-band \textbf{k}$\cdot$\textbf{p} calculations.
  \emph{Semicond.~Sci.~Technol.} \textbf{2015}, \emph{30}, 094009\relax
\mciteBstWouldAddEndPuncttrue
\mciteSetBstMidEndSepPunct{\mcitedefaultmidpunct}
{\mcitedefaultendpunct}{\mcitedefaultseppunct}\relax
\EndOfBibitem
\bibitem[Usman and O'Reilly(2014)Usman, and O'Reilly]{Usman_APL_2014}
Usman,~M.; O'Reilly,~E.~P. Atomistic tight-binding study of electronic
  structure and interband optical transitions in GaBiAs/GaAs quantum wells.
  \emph{Appl. Phys. Lett.} \textbf{2014}, \emph{104}, 071103\relax
\mciteBstWouldAddEndPuncttrue
\mciteSetBstMidEndSepPunct{\mcitedefaultmidpunct}
{\mcitedefaultendpunct}{\mcitedefaultseppunct}\relax
\EndOfBibitem
\bibitem[Klimeck et~al.(2007)Klimeck, Ahmed, Kharche, Korkusinski, Usman,
  Prada, and Boykin]{Klimeck_IEEETED_2007_2}
Klimeck,~G.; Ahmed,~S.~S.; Kharche,~N.; Korkusinski,~M.; Usman,~M.; Prada,~M.;
  Boykin,~T. Atomistic simulation of realistically sized nanodevices using
  {NEMO 3-D} -- part {I}: applications. \emph{IEEE Trans.~Electron.~Dev.}
  \textbf{2007}, \emph{54}, 2090\relax
\mciteBstWouldAddEndPuncttrue
\mciteSetBstMidEndSepPunct{\mcitedefaultmidpunct}
{\mcitedefaultendpunct}{\mcitedefaultseppunct}\relax
\EndOfBibitem
\bibitem[Ahmed et~al.(2009)Ahmed, Kharche, Rahman, Usman, Lee, Ryu, Bae, Clark,
  Haley, Naumov, Saied, Korkusinski, Kennel, McLennan, Boykin, and
  Klimeck]{Ahmed2009}
Ahmed,~S. et~al.  In \emph{Encyclopedia of Complexity and Systems Science};
  Meyers,~R.~A., Ed.; Springer New York: New York, NY, 2009; pp
  5745--5783\relax
\mciteBstWouldAddEndPuncttrue
\mciteSetBstMidEndSepPunct{\mcitedefaultmidpunct}
{\mcitedefaultendpunct}{\mcitedefaultseppunct}\relax
\EndOfBibitem
\bibitem[Qian et~al.(2008)Qian, Li, Gradecak, Park, Dong, Ding, Wang, and
  Lieber]{Qian_NM_2008}
Qian,~F.; Li,~Y.; Gradecak,~S.; Park,~H.-G.; Dong,~Y.; Ding,~Y.; Wang,~Z.~L.;
  Lieber,~C. {Multi-quantum-well nanowire heterostructures for
  wavelength-controlled lasers}. \emph{Nature Materials} \textbf{2008},
  \emph{7}, 701\relax
\mciteBstWouldAddEndPuncttrue
\mciteSetBstMidEndSepPunct{\mcitedefaultmidpunct}
{\mcitedefaultendpunct}{\mcitedefaultseppunct}\relax
\EndOfBibitem
\bibitem[Usman et~al.(2009)Usman, Ryu, Woo, Ebert, and
  Klimeck]{Usman_IEEE_2009}
Usman,~M.; Ryu,~H.; Woo,~I.; Ebert,~D.; Klimeck,~G. {Moving toward nano-TCAD
  through multimillion-atom quantum-dot simulations matching experimental
  data}. \emph{IEEE Trans. Nanotech.} \textbf{2009}, \emph{8}, 330\relax
\mciteBstWouldAddEndPuncttrue
\mciteSetBstMidEndSepPunct{\mcitedefaultmidpunct}
{\mcitedefaultendpunct}{\mcitedefaultseppunct}\relax
\EndOfBibitem
\bibitem[Keating(1966)]{Keating_PR_1966}
Keating,~P.~N. Effect of invariance requirements on the elastic strain energy
  of crystals with application to the diamond structure. \emph{Phys. Rev.}
  \textbf{1966}, \emph{145}, 637\relax
\mciteBstWouldAddEndPuncttrue
\mciteSetBstMidEndSepPunct{\mcitedefaultmidpunct}
{\mcitedefaultendpunct}{\mcitedefaultseppunct}\relax
\EndOfBibitem
\bibitem[Lee et~al.(2004)Lee, Lazarenkova, von Allmen, Oyafuso, and
  Klimeck]{Lee_1}
Lee,~S.; Lazarenkova,~O.~L.; von Allmen,~P.; Oyafuso,~F.; Klimeck,~G. Effect of
  wetting layers on the strain and electronic structure of InAs self-assembled
  quantum dots. \emph{Phys. Rev. B} \textbf{2004}, \emph{70}, 125307\relax
\mciteBstWouldAddEndPuncttrue
\mciteSetBstMidEndSepPunct{\mcitedefaultmidpunct}
{\mcitedefaultendpunct}{\mcitedefaultseppunct}\relax
\EndOfBibitem
\bibitem[Singh(1985)]{Singh_JAP_1985}
Singh,~J. {Role of interface roughness and alloy disorder in photoluminescence
  in quantum-well structures}. \emph{J. Appl. Phys.} \textbf{1985}, \emph{57},
  5433\relax
\mciteBstWouldAddEndPuncttrue
\mciteSetBstMidEndSepPunct{\mcitedefaultmidpunct}
{\mcitedefaultendpunct}{\mcitedefaultseppunct}\relax
\EndOfBibitem
\bibitem[Usman et~al.(2011)Usman, Heck, Clarke, Spencer, Ryu, Murray, and
  Klimeck]{Usman_JAP_2011_2}
Usman,~M.; Heck,~S.; Clarke,~E.; Spencer,~P.; Ryu,~H.; Murray,~R.; Klimeck,~G.
  {Experimental and theoretical study of polarization-dependent optical
  transitions in InAs quantum dots at telecommunication-wavelengths (1300-1500
  nm)}. \emph{J. Appl. Phys.} \textbf{2011}, \emph{109}, 104510\relax
\mciteBstWouldAddEndPuncttrue
\mciteSetBstMidEndSepPunct{\mcitedefaultmidpunct}
{\mcitedefaultendpunct}{\mcitedefaultseppunct}\relax
\EndOfBibitem
\bibitem[Usman et~al.(2011)Usman, Inoue, Harda, Klimeck, and
  Kita]{Usman_PRB2_2011}
Usman,~M.; Inoue,~T.; Harda,~Y.; Klimeck,~G.; Kita,~T. Experimental and
  atomistic theoretical study of degree of polarization from multilayer
  InAs/GaAs quantum dot stacks. \emph{Phys. Rev. B} \textbf{2011}, \emph{84},
  115321\relax
\mciteBstWouldAddEndPuncttrue
\mciteSetBstMidEndSepPunct{\mcitedefaultmidpunct}
{\mcitedefaultendpunct}{\mcitedefaultseppunct}\relax
\EndOfBibitem
\bibitem[Usman(2018)]{Usman_PRM_2018}
Usman,~M. Large-scale atomistic simulations demonstrate dominant alloy disorder
  effects in GaBiAs/GaAs multiple quantum wells. \emph{Phys. Rev. Materials}
  \textbf{2018}, \emph{2}, 044602\relax
\mciteBstWouldAddEndPuncttrue
\mciteSetBstMidEndSepPunct{\mcitedefaultmidpunct}
{\mcitedefaultendpunct}{\mcitedefaultseppunct}\relax
\EndOfBibitem
\bibitem[Usman(2012)]{Usman_PRB_2012}
Usman,~M. {Atomistic theoretical study of electronic and polarization
  properties of single and vertically stacked elliptical InAs quantum dots}.
  \emph{Phys. Rev. B} \textbf{2012}, \emph{86}, 155444\relax
\mciteBstWouldAddEndPuncttrue
\mciteSetBstMidEndSepPunct{\mcitedefaultmidpunct}
{\mcitedefaultendpunct}{\mcitedefaultseppunct}\relax
\EndOfBibitem
\bibitem[Greif et~al.(2018)Greif, Jagsch, Wagner, and
  Schliwa]{Greif_ACSPhotonics_2018}
Greif,~L.; Jagsch,~S.; Wagner,~M.; Schliwa,~A. Tuning the Emission
  Directionality of Stacked Quantum Dots. \emph{ACS Photonics} \textbf{2018},
  \emph{5}, 12\relax
\mciteBstWouldAddEndPuncttrue
\mciteSetBstMidEndSepPunct{\mcitedefaultmidpunct}
{\mcitedefaultendpunct}{\mcitedefaultseppunct}\relax
\EndOfBibitem
\bibitem[Inoue et~al.(2010)Inoue, Asada, Yasuoka, Kojima, Kita, and
  Wada]{Inoue_APL_2010}
Inoue,~T.; Asada,~M.; Yasuoka,~N.; Kojima,~O.; Kita,~T.; Wada,~O. {}.
  \emph{Appl. Phys. Lett.} \textbf{2010}, \emph{96}, 211906\relax
\mciteBstWouldAddEndPuncttrue
\mciteSetBstMidEndSepPunct{\mcitedefaultmidpunct}
{\mcitedefaultendpunct}{\mcitedefaultseppunct}\relax
\EndOfBibitem
\bibitem[Wang et~al.(2001)Wang, Gudiksen, Duan, Cui, and
  Lieber]{Wang_Science_2001}
Wang,~J.; Gudiksen,~M.; Duan,~X.; Cui,~Y.; Lieber,~C. {Highly Polarized
  Photoluminescence and Photodetection from Single Indium Phosphide Nanowires}.
  \emph{Phys. Rev. B} \textbf{2001}, \emph{293}, 1455\relax
\mciteBstWouldAddEndPuncttrue
\mciteSetBstMidEndSepPunct{\mcitedefaultmidpunct}
{\mcitedefaultendpunct}{\mcitedefaultseppunct}\relax
\EndOfBibitem
\bibitem[Casadei et~al.(2015)Casadei, Llado, Amaduzzi, Russo-Averchi, Ruffer,
  Heiss, Negro, and Morral]{Casadei_SciRep_2015}
Casadei,~A.; Llado,~E.~A.; Amaduzzi,~F.; Russo-Averchi,~E.; Ruffer,~D.;
  Heiss,~M.; Negro,~L.~D.; Morral,~A.~F. {Polarization response of nanowires}.
  \emph{Sci. Rep.} \textbf{2015}, \emph{5}, 7651\relax
\mciteBstWouldAddEndPuncttrue
\mciteSetBstMidEndSepPunct{\mcitedefaultmidpunct}
{\mcitedefaultendpunct}{\mcitedefaultseppunct}\relax
\EndOfBibitem
\bibitem[Hadar et~al.(2013)Hadar, Hitin, Sitt, Faust, and
  Banin]{Hadar_JPCL_2013}
Hadar,~I.; Hitin,~G.; Sitt,~A.; Faust,~A.; Banin,~U. {Polarization Properties
  of Semiconductor Nanorod Heterostructures: From Single Particles to the
  Ensemble}. \emph{J. Phys. Chem. Lett.} \textbf{2013}, \emph{3}, 502\relax
\mciteBstWouldAddEndPuncttrue
\mciteSetBstMidEndSepPunct{\mcitedefaultmidpunct}
{\mcitedefaultendpunct}{\mcitedefaultseppunct}\relax
\EndOfBibitem
\bibitem[Zhi et~al.(2016)Zhi, Tao, Liu, Zhuang, Dai, Li, Zhang, Xie, Chen, and
  Zhang]{Zhi_IEEEP_2016}
Zhi,~T.; Tao,~T.; Liu,~B.; Zhuang,~Z.; Dai,~J.; Li,~Y.; Zhang,~G.; Xie,~Z.;
  Chen,~P.; Zhang,~R. {Polarized Emission From InGaN/GaN Single Nanorod
  Light-Emitting Diode}. \emph{IEEE Photonics Technology Letters}
  \textbf{2016}, \emph{28}, 721\relax
\mciteBstWouldAddEndPuncttrue
\mciteSetBstMidEndSepPunct{\mcitedefaultmidpunct}
{\mcitedefaultendpunct}{\mcitedefaultseppunct}\relax
\EndOfBibitem
\bibitem[Usman(2011)]{Usman_JAP_2011}
Usman,~M. {In-plane polarization anisotropy of ground state optical intensity
  in InAs/GaAs quantum dots}. \emph{J. Appl. Phys.} \textbf{2011}, \emph{110},
  094512\relax
\mciteBstWouldAddEndPuncttrue
\mciteSetBstMidEndSepPunct{\mcitedefaultmidpunct}
{\mcitedefaultendpunct}{\mcitedefaultseppunct}\relax
\EndOfBibitem
\bibitem[Broderick et~al.(2013)Broderick, Usman, and
  O'Reilly]{Broderick_PSSB_2013}
Broderick,~C.; Usman,~M.; O'Reilly,~E. {12-band k.p model for dilute bismide
  alloys of (In)GaAs derived from supercell calculations}. \emph{Physica Status
  Solidi (b)} \textbf{2013}, \emph{250}, 733\relax
\mciteBstWouldAddEndPuncttrue
\mciteSetBstMidEndSepPunct{\mcitedefaultmidpunct}
{\mcitedefaultendpunct}{\mcitedefaultseppunct}\relax
\EndOfBibitem
\bibitem[Rajpalke et~al.(2013)Rajpalke, Linhart, Birkett, Yu, Scanlon,
  Buckeridge, Jones, Ashwin, and Veal]{Rajpalke_APL_2013}
Rajpalke,~M.~K.; Linhart,~W.~M.; Birkett,~M.; Yu,~K.~M.; Scanlon,~D.~O.;
  Buckeridge,~J.; Jones,~T.~S.; Ashwin,~M.~J.; Veal,~T.~D. {Growth and
  properties of GaSbBi alloys}. \emph{Appl. Phys. Lett.} \textbf{2013},
  \emph{103}, 142106\relax
\mciteBstWouldAddEndPuncttrue
\mciteSetBstMidEndSepPunct{\mcitedefaultmidpunct}
{\mcitedefaultendpunct}{\mcitedefaultseppunct}\relax
\EndOfBibitem
\bibitem[Bushell et~al.(2019)Bushell, Broderick, Nattermann, Joseph, Keddie,
  Rorison, Volz, and Sweeney]{Bushell_SR_2019}
Bushell,~Z.; Broderick,~C.; Nattermann,~L.; Joseph,~R.; Keddie,~J.;
  Rorison,~J.; Volz,~K.; Sweeney,~S. {Giant bowing of the band gap and
  spin-orbit splitting energy in GaP$_{1−x}$Bi$_x$ dilute bismide alloys}.
  \emph{Scientific Reports} \textbf{2019}, \emph{9}, 6835\relax
\mciteBstWouldAddEndPuncttrue
\mciteSetBstMidEndSepPunct{\mcitedefaultmidpunct}
{\mcitedefaultendpunct}{\mcitedefaultseppunct}\relax
\EndOfBibitem
\end{mcitethebibliography}

\providecommand*\mcitethebibliography{\thebibliography}
\csname @ifundefined\endcsname{endmcitethebibliography}
  {\let\endmcitethebibliography\endthebibliography}{}

\end{document}